\documentclass[12pt,preprint]{aastex}
\shorttitle{Observations of Non-thermal Velocity and Comparison with Alfv\'en Wave Turbulence Model in Solar Active Regions}
\shortauthors{}
\usepackage{booktabs} % for better table rules
\graphicspath{{./}{figures/}}
\usepackage{graphicx}
\usepackage{subfigure}
\usepackage{placeins}
\usepackage{hhline}
\usepackage{media9}

\author{M. Asgari-Targhi\altaffilmark{1}, D. H. Brooks\altaffilmark{2,3}, M. Hahn\altaffilmark{4},
S. Imada\altaffilmark{5}, E. Tajfirouze\altaffilmark{6}, and D. W. Savin\altaffilmark{4}}
\altaffiltext{1}{Harvard-Smithsonian Center for Astrophysics, 60 Garden Street MS-15,
Cambridge, MA 02138, USA}
\altaffiltext{2}{Department of Physics \& Astronomy, George Mason University, 4400 University Drive,
Fairfax, VA 22030, USA}
\altaffiltext{3}{Present address: Hinode Team, ISAS/JAXA, 3-1-1
Yoshinodai, Chuo-ku, Sagamihara, Kanagawa 252-5210, Japan}
\altaffiltext{4}{Columbia Astrophysics Laboratory, Columbia University, 550 West 120th Street, New York, NY 10027, USA}
\altaffiltext{5}{Solar-Terrestrial Environment Laboratory, Nagoya University,
Furo-cho, Chikusa-ku, Nagoya, Aichi, 464-8601, Japan}
\altaffiltext{6}{Mathematics, Physics and Electrical Engineering, Northumbria University, Newcastle Upon Tyne NE1 8ST, UK}

\begin{document}

\title{Observations of Non-thermal Velocities and Comparisons with an Alfv\'en Wave Turbulence Model in Solar Active Regions}

\begin{abstract}
We present a study of spectral line width measurements from 
the Extreme Ultraviolet Imaging Spectrometer (EIS) on {\it Hinode}. We used spectral line profiles of Fe {\sc xvi} 262.984 {\AA}, Fe {\sc xiv} 264.787 {\AA}, Fe {\sc xiv} 270.519 {\AA}, Fe {\sc xiv} 274.203 {\AA}, and Fe {\sc xv} 284.160 {\AA}, and studied 11 active regions. Previous studies of spectral line widths 
have shown that in hot loops in the cores of active regions, the observed non-thermal velocities are smaller than predicted from models of reconnection jets in the corona or shock heating associated with Alfv\'{e}n waves. The observed line widths are also inconsistent with models of chromospheric evaporation due to coronal nanoflares. We show that recent advances in higher 
resolution Alfv\'{e}n wave turbulence modeling enables us to obtain non-thermal velocities similar to those measured in 
 active regions. The observed non-thermal velocities for the 11 active regions in our study are in the range of 17$-$30 $\rm km ~ s^{-1}$, consistent with the spectral line non-thermal widths predicted from our model of 16 interacting flux tubes, which are in the range of ~15$-$37  $\rm km ~ s^{-1}$. 
  \end{abstract}


\section{Introduction}

Solar active regions possess closed magnetic field lines that are rooted in the
Sun at both ends and are known as coronal loops. Some of these loops have temperatures in the range of 1$-$3 MK.
The physical processes by which these loops are heated are not yet fully understood. However, 
it is generally assumed that the plasma is heated by dissipation of magnetic disturbances
that propagate into the corona from the convection zone below the photosphere. The two main models 
are nanoflare-heating and wave-heating \citep[e.g.,][]{Ionson1985, Milano1997, Mandrini2000}. In the nanoflare-heating model, 
the convective flows below the photosphere are assumed to cause twisting and
braiding of the coronal magnetic field lines. The magnetic free energy associated with
the braided fields is released in brief reconnection events (``nanoflares") that occur
throughout the corona \citep[][]{Parker1972, Parker1983, Parker1988, Cargill2004,
Patsourakos2006, Patsourakos2009, Imada2012}. In wave-heating models, the
convective flows are thought to produce magnetohydrodynamic (MHD) waves such as 
Alfv\'{e}n and kink waves that are transverse to the background magnetic field. These waves propagate upward along the magnetic field and dissipate
their energy in the corona \citep[][] {Alfven1947, Coleman1968, Uchida1974, Wentzel1974,
Hollweg1982, Heyvaerts1983, Antolin2010, Matsumoto2010, McIntosh2011, vanB2011}. 

Observations from space- and ground-based telescopes are necessary to test and evaluate these 
heating models. For example, there is evidence for footpoint motions that can either create twists within the field lines 
or generate MHD waves. Observational constraints on the amplitudes of the magnetic and velocity perturbations 
in the corona are essential to understand how the energy is transferred from the lower atmosphere into the corona. 
Any proposed heating mechanism is expected to provide  a self-consistent picture of the chromosphere and corona and 
to generate sufficient heating of the coronal plasma to the observed temperatures. 

The Doppler widths of coronal emission lines provide significant constraints on unresolved
plasma flows in the solar atmosphere.  The spectral line width mainly consists of three parts, the thermal width, the instrumental width, and the non-thermal width. 
The thermal line width is usually interpreted under the assumption that the ion temperature is equivalent to the electron temperature. 
The non-thermal line width is associated with flows caused by turbulence or magnetic reconnection \citep[][]{Hara1999, dos2007, ima2008}. 
The Extreme Ultraviolet (EUV) Imaging Spectrometer (EIS) on
the {\it Hinode} satellite has provided information on spectral line broadening for individual coronal loops
in active regions \citep[e.g.,][]{Young2007, Tripathi2009, Warren2011a, Tripathi2011}.
There have also been ground-based observations of spectral line broadening \citep[e.g.,][]
{Ichimoto1995, Hara1999}, and satellite-based measurements from X-ray lines \citep[][]{Acton1981}. The line
broadening is generally attributed to MHD waves and/or turbulent flows \citep[][]{Doschek1977,
Tian2011, Doschek2012}, but an alternative interpretation is that the
excess broadening is due to ion temperatures being larger than electron temperatures
\citep[][]{Billings1962, Knight1974, Imada2009}.  

Alfv\'{e}n waves have been hypothesized to play a major role in heating the solar corona.
To put constraints on the amplitudes of Alfv\'{e}n waves in an active region
observed above the solar limb, \citet[][]{Hara1999} adopted spectroscopic observations with a coronagraph at the Norikura Solar
Observatory \citep[also see][]{Ichimoto1995}. They showed that the non-thermal
velocities for Fe~{\sc x} 6374 {\AA}, Fe~{\sc xiv} 5303 {\AA}, and Ca~{\sc xv} 5694 {\AA} have ranges of 14$-$20, 10$-$18, and 
16$-$26 $\rm km ~ s^{-1}$, respectively. They also examined the
relationship between the widths of the emission lines and the orientation of the coronal loops
relative to the line of sight (LOS). They found that for face-on loops, the non-thermal velocity
is approximately constant along the loop; while for edge-on loops, there is a decrease in
velocity of about 3$-$5 $\rm km ~ s^{-1}$ towards the loop top. This decrease in velocity was
interpreted as evidence for the existence of Alfv\'{e}n waves in coronal loops. At the loop
top for edge-on loops the magnetic field is parallel to the LOS, so the transverse waves do not
contribute to the LOS velocity. 

Based on these observations, \citet[][]{Hara1999}
concluded that Alfv\'{e}n waves might have velocity amplitudes of 3$-$5 $\rm km ~ s^{-1}$.
However, we suggest that these values may represent a lower limit on the Alfv\'{e}n wave amplitude.
In observations of face-on loops, \citet[][]{Hara1999} observed non-thermal velocities in the range of 
10$-$18 $\rm km ~ s^{-1}$, which represents motions perpendicular to the background field.
It seems likely to us that all such perpendicular motions are associated with Alfv\'{e}nic waves \citep[][]{Asgari2014, vanB2017}.
These waves could either originate in the photosphere or be produced by reconnection events in the
corona. In either case, Alfv\'{e}n and/or kink waves are the only plausible candidates for
producing transverse motions in the corona. Therefore, we suggest that the observations by
\citet[][]{Hara1999} are consistent with Alfv\'{e}n wave amplitudes in the range 10 - 18
$\rm km ~ s^{-1}$.

 To place
observational constraints on wave-heating theories, we must determine the contribution of the
Alfv\'{e}nic waves to the observed non-thermal velocities. In a series of papers, we have studied the dynamics of Alfv\'{e}n waves in coronal loops, using three-dimensional (3D) reduced MHD 
models \citep[RMHD;][]{vanB2011, Asgari2012, Asgari2013, vanB2014, vanB2017}. We showed the wave energy flux from the Alfv\'{e}n wave turbulence is 
sufficient to heat the coronal loops to a temperature of about 2.5 MK. Alfv\'{e}n wave turbulence model also predicts root-mean-square (RMS) velocity amplitudes of the Alfv\'{e}n waves to be in the range 15$-$37 $\rm km ~ s^{-1}$ in the corona, with the highest velocities generally occurring near the loop top \citep[][]{Asgari2014}. 

A key test of this model is to compare the predicted velocities with observations of non-thermal line widths of coronal emission
lines. \citet[][]{Asgari2014} made such a comparison, using EIS data.
We compared the observed non-thermal line broadening in Fe {\sc xii} 192.394 {\AA} for individual coronal loops in a single active region of 2012 September 
7 with the LOS velocity from our Alfv\'{e}n wave turbulence model of the loops. We found that footpoint velocities in the range 0.30$-$1.50 $\rm km ~ s^{-1}$ can reproduce the observed coronal non-thermal widths of 15$-$37 $\rm km ~ s^{-1}$.  
The footpoint velocities, one of the input parameters of our model are consistent with observed motions of magnetic elements in the photosphere \citep[][]{Abramenko2011, Chitta2012}.  Furthermore, we found that in order to produce the observed non-thermal velocities, we needed to introduce a random flow component parallel to the magnetic field, in addition to the perpendicular velocity due to Alfv\'{e}n waves.

The analysis of \citet[][]{Asgari2014}  used the $2^{\prime \prime}$ slit of 
EIS, but assumed an instrumental line width appropriate for a $1^{\prime \prime}$ slit. Therefore, the instrument width was underestimated, which led to an overestimation of the EIS non-thermal velocity. Subsequently, \citet[][]{Brooks2016}, performed a systematic analysis of the instrumental effects that influence EIS spectral line width measurements. Here, we follow their analysis to improve our line width measurements and compare the EIS observations with our Alfv\'{e}n wave turbulence model. 

The objective of this paper is twofold. First,  we present a systematic observational study of non-thermal velocity in 11 active regions, only one of which was analyzed in \citet{Asgari2014}. Second, we use an improved Alfv\'en wave turbulence model containing multiple flux tubes with high spatial resolution \citep[][]{vanB2017} and compare the simulated non-thermal velocities from our multiple flux tube model with measurements averaged over the whole loop arcade of an active region. This is in contrast to \citet{Asgari2014}, where the dynamics of flows within individual loops were analyzed using a reduced MHD approximation of a single flux tube and compared with non-thermal velocities derived from EIS observations for individual loops. The emission of the spectral lines we use, from Fe {\sc xiv}, {\sc xv}, and {\sc xvi}, represents an integration along the LOS and therefore potentially contains contributions from both the active region loop arcade, and unresolved foreground and background emission.
\citet[][]{Hahn2023} have quantified the unresolved component contribution to these lines in an active region, indicating substantial contributions, with an increasing impact as the line formation temperature decreases. This contribution may also be active region dependent \citep[][]{Brown2008}. Considering the complexity of the emitting structures, we expect the bright arcade and unresolved (potentially cooler) background to represent multiple loop structures; so these are the appropriate observations to compare with our multi-loop model. 
 
The rest of this paper is organized as follows. In Section 2, we discuss the observations and issues involved in measuring non-thermal velocities, including the uncertainties in the instrumental contribution to the line width. In Section 3, we discuss the results from the Alfv\'{e}n wave turbulence models. 
Section 4 presents the effects of Alfv\'{e}n wave turbulence in multiple flux tubes on spectral line broadening. Finally, the discussion and conclusion are given in Section 5. 

\section{Observations}

\subsection{Non-thermal Line Widths and Instrumental Width}

The observed line width is characterized by its Full Width  at Half Maximum (FWHM) computed from: 
\begin{equation}
{\rm FWHM} = \left[\left ( 1.665 \frac{\lambda} {c} \right)^{2} \left(
  \frac{2k_{\rm B}T} {M} + W^2 \right)  + (\Delta \lambda)^2 \right]^{1/2},
\label{eq:Specwidth}
\end{equation}
where $\lambda$ is the wavelength, $c$ is the speed of light,
$k_{\rm B}$ is the Boltzmann constant, $T$ is the ion temperature, 
$M$ is the ion mass, $W$ is the $1/e$ half width of the non-thermal velocity distribution, and $\Delta \lambda$ is the instrumental width. In this formula all three 
components are considered to have Gaussian distributions.
The instrumental width profile is crucial for computing the non-thermal width. 
 
For the observational portion of our study, we use the EIS instrument.  EIS observes the Sun in two spectral passbands, detecting 
spectral lines that emanate from a series of ionization states. In specific, it observes the solar corona and upper transition region with high spectral and spatial resolution over
170$-$210 and 250$-$290 {\AA} \citep[][]{Korendyke2006, Culhane2007}.
The line centroids and profile widths enable one to measure the motions of 
plasmas and turbulent or non-thermal line broadenings. In the EIS data, the $\Delta \lambda$ instrumental 
width is similar in magnitude to $W$ and the thermal broadening. We account for this $\Delta \lambda$ using the calibration tabulated by \citet{Young2011}. 

\subsection{Observational Results}

In this study, we considered 11 active regions. The datasets we used are listed in Table 1.  The first three were observed with the $1^{\prime \prime}$ resolution slit, and the rest were observed with the $2^{\prime \prime}$ resolution slit.
We included data obtained using the two different slit resolutions to determine if the slit resolution has any
significant influence on the measurements of the total line broadening. Our analysis did not find any systematic differences.
We used spectral line profiles of Fe {\sc xvi} 262.984 {\AA}, Fe {\sc xiv} 264.787 {\AA}, Fe  {\sc xiv} 270.519  {\AA}, Fe {\sc xiv} 274.203  {\AA}, and Fe  {\sc xv} 284.160  {\AA}. 

\begin{table}[!]
\setlength{\abovecaptionskip}{10pt} % Adjust the spacing between table number and caption
\centering
%\resizebox{\textwidth}{!}{%
%\caption{Observations of Active Regions}
%\captionsetup{justification=centering}
\caption{Observations of Active Regions}
\begin{tabular}{|c|c|c|c|}
\hline
 NOAA Number  & EIS Data Set  & Slit Size \\
 \hline
11029 &${\rm{eis}\_{l0}\_{20091026} \_{035809} }$  &1$^{\prime\prime}$ \\ 
11247 & ${\rm{eis}\_{l0}\_{20110713}\_{050720}}$ & 1$^{\prime\prime}$  \\
11250 & ${\rm{eis}\_{l0}\_{20110715}\_{094619}}$ & 1$^{\prime\prime}$  \\
11082 & $\rm{eis}\_l0\_20100621\_142401 $ & 2$^{\prime\prime}$  \\
11087 & $\rm{eis}\_l0\_20100716\_001932 $& 2$^{\prime\prime}$  \\
11093 & $\rm{eis}\_l0\_20100810\_223844 $& 2$^{\prime\prime}$ \\
11108 & $\rm{eis}\_l0\_20100922\_112633 $ & 2$^{\prime\prime}$ \\
11117 & $\rm{eis}\_l0\_20101026\_104913 $ & 2$^{\prime\prime}$ \\
11127 &$\rm{eis}\_l0\_20101123\_221756$ & 2$^{\prime\prime}$   \\
11135& $\rm{eis}\_l0\_20101217\_155443 $& 2$^{\prime\prime}$ \\
11564& $\rm{eis}\_l0\_20120907\_070549 $& 2$^{\prime\prime}$ \\
\hline
\end{tabular}
\label{tab1}
\end{table}

We used the standard routine eis\_prep, which is available within the EIS branch of \textit{solarsoft}, to 
correct for instrumental effects such as the dark current pedestal, and contamination by dusty, warm, and hot pixels \citep[][]{Brooks2016}. 
The routine also corrects for the movement of the spectrum on the CCDs due to thermal orbital effects using a neural
network model \citep[][]{Kamio2010}. We did not apply the absolute calibration because this has a tendency to 
increase the line widths. Recent studies highlight the importance of accurate measurements of the instrumental line 
width, together with detailed assessment of other instrumental characteristics, in deducing
the relevant non-thermal velocity \cite[][]{Brooks2016, Testa2016}. The method we used is the same as \citet[][]{Brooks2016}.
We fit a single Gaussian function to the spectral line profiles of all lines except for Fe {\sc xiv} 270.520 {\AA}
where we also fit the nearby Mg {\sc vi} 270.394 {\AA} line to improve the line fit. 
From the fits, we determined the line intensity, Doppler velocity, and line width.

When EIS raster images were missing data, we used images from the Atmospheric Imaging Assembly \citep[AIA;][]{Lemen2012} for context. AIA is a multi-layer telescope on-board \textit{Solar Dynamic Observatory} ({\it SDO}). It provides full disk images of the Sun at high spatial resolution ($0.6^{\prime \prime}$ pixels) and high cadence ($ \sim 12~{\rm s}$) in the EUV passbands. We used the standard processing (aia\_prep routine) to obtain the level-1.0 data (bad pixels removal, despiking, and flat fielding) from the level-0 data. We converted the pixel image to arc second.  We extracted regions with EIS dimensions from the full disc images of \textit{SDO}/AIA. The EIS $1^{\prime \prime} $ slit scans an area of  $128^{\prime \prime}\times 128^{\prime \prime}$ and the $2^{\prime \prime}$ slit rasters over an area of $120^{\prime \prime}\times 160^{\prime \prime}$. 
 
\begin{figure}[http]
\epsscale{1.00}
\plotone{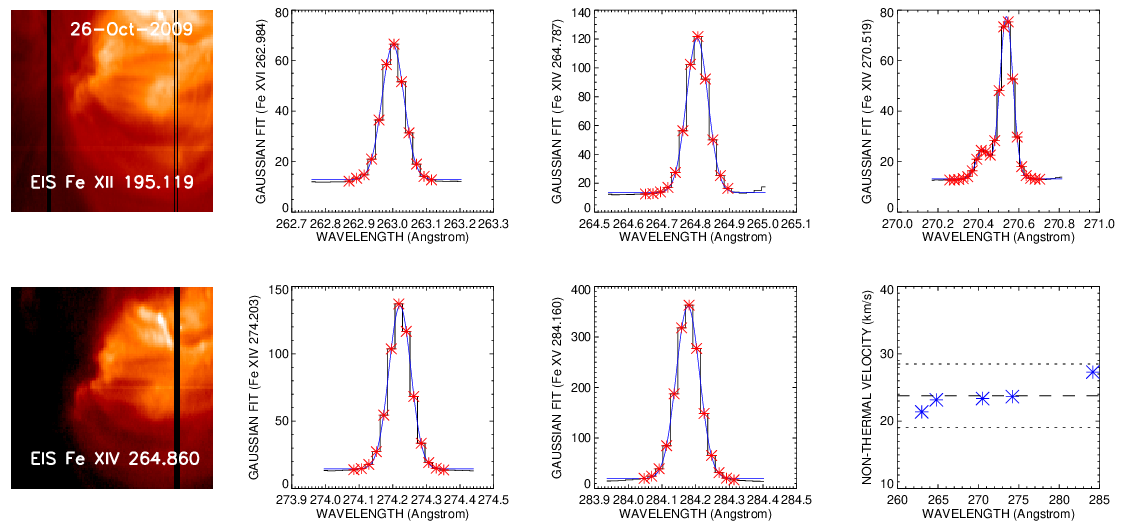}
\caption{Active region NOAA 11029 observed by EIS with $1^{\prime \prime}$ resolution slit on 2009 October 26 in Fe {\sc xii} 195.12 {\AA} and Fe {\sc xiv} 264.860 {\AA} (left column). The next five panels show the Gaussian fits to the spectral lines Fe {\sc xvi} 262.984 {\AA}, Fe {\sc xiv} 264.787 {\AA}, Fe {\sc xiv} 270.519 {\AA}, Fe {\sc xiv} 274.203 {\AA}, and Fe {\sc xv} 284.160 {\AA}. The unit for the y-axis is the uncalibrated data number (DN). The histogram and red asterisks show the data and the smooth blue curve, the fit. The lower right panel shows the non-thermal velocity derived from the Gaussian fits for all the lines (blue asterisks). The dashed line shows the mean for this active region, and the dotted lines show the approximate variation calculated from all the measurements in all the active regions.}
\label{fig:fig1}
\end{figure}

An example of the Gaussian fits for one of the active regions studied is shown in Figure~\ref{fig:fig1}. The line profiles are averages over the whole active region. The results for the other active regions are summarized in Appendix A. The first column in Figure~\ref{fig:fig1} shows the EIS  intensity maps for the active region NOAA 11029 observed on  2009 October 26 in two 
EUV bands, 195.118  {\AA} and 264.860 {\AA}. These channels contain strong lines of Fe {\sc xii} and Fe {\sc xiv} and are 
sensitive to plasma emitted at temperatures $\sim$  $\log T =6.1$ and 6.3 respectively, where $T$ is measured in units of K.
From the line fits, we find that as the wavelength increases from 262.984 {\AA} to 284.160 {\AA}, there is a change in the non-thermal velocity from 21 $\rm km ~ s^{-1}$, reaching a value of 27 $\rm km ~ s^{-1}$. This change could be an indication of a temperature effect in these active regions. Alternatively, it could be the result of a wavelength dependent instrumental effect. The higher value of the non-thermal velocity for Fe {\sc xv} 284.160 {\AA}, however, could be due to the contamination of this line with Al {\sc ix} 284.015 {\AA} \citep[][]{Brooks2016}. The consistency of the values for the two Fe {\sc xiv} lines suggests that the observed increasing non-thermal velocity might be a temperature dependence. However, we have estimated the uncertainty in the measurements by calculating the standard deviation in non-thermal velocities throughout all the datasets we analyze and find a value of 4.7\,km s$^{-1}$ (shown by the dotted lines in the figure). The variation with temperature, therefore, is within the measurement uncertainties.

\begin{figure}[http]
\epsscale{0.70}
%\plotone{f2.pdf}
\plotone{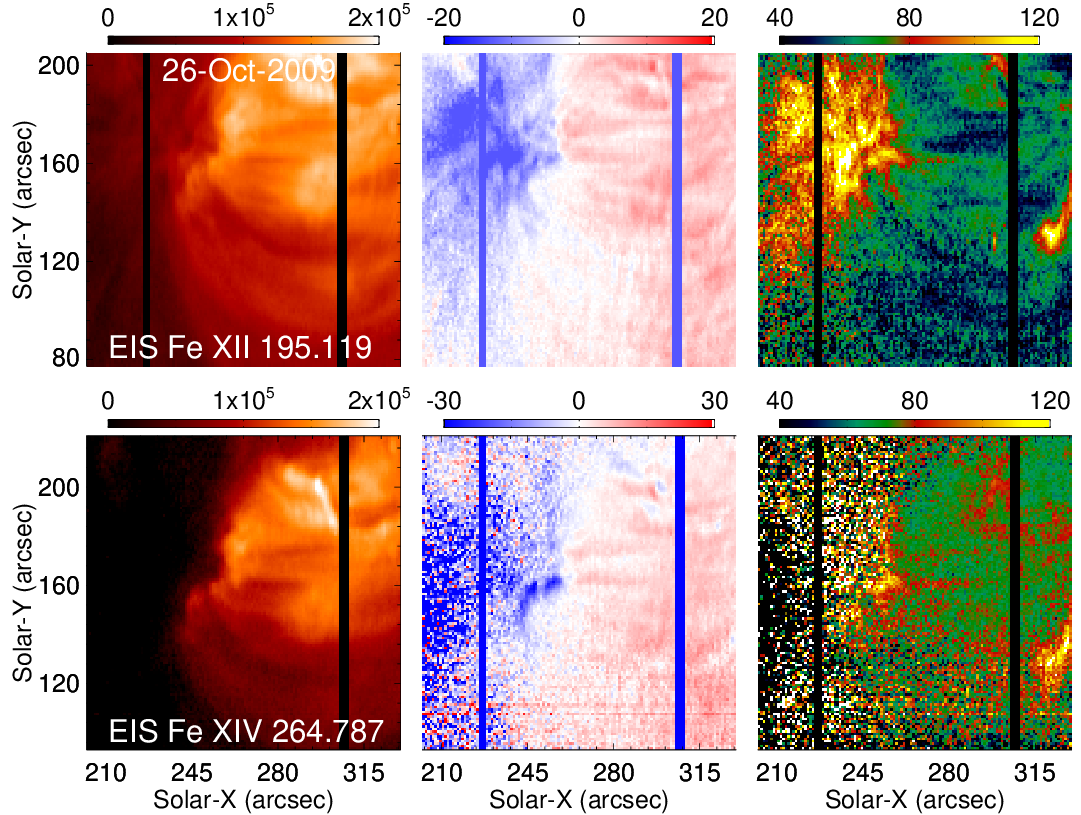}
\caption{Imaging spectra of intensity in units of $\rm  ergs~\rm{cm}^{-2}~s^{-1}~{\rm steradian}^{-1}$, Doppler velocity in units of $\rm km ~ s^{-1}$ and non-thermal velocity (FWHM) in units of $\rm km ~ s^{-1}$ for the active region NOAA 11029 in Fe {\sc xii} 195.12 {\AA} (top row) and Fe {\sc xiv} 264.787 {\AA} (bottom row).}
\label{fig:fig2}
\end{figure}

Figure~\ref{fig:fig2} shows the same active region NOAA 11029 in Fe  {\sc xii} 195.118 {\AA} and Fe  {\sc xiv} 264.787 {\AA}. For the Fe {\sc xii} 195.118 {\AA} line we removed the contribution of the blended weak line (Fe {\sc xii} 195.18 {\AA}). Similar EIS maps for the rest of the active regions listed in Table 1 are shown in Appendix A. For the 11 active regions we studied, we find non-thermal velocities in the range of 17--30\,km s$^{-1}$.

\clearpage
\section{Alfv\'{e}n Wave Turbulence Dynamics in Multiple Coronal Loops}

We have previously studied the dynamics of Alfv\'{e}n waves in a single flux tube representing a solar coronal loop in a series of papers using 3D RMHD models \citep[][]{vanB2011, Asgari2012, Asgari2013, vanB2014, vanB2017}. In these models, the waves are generated by the foot-point motions at the photosphere. The random footpoint motions in our model are created by granule-scale convective flows with velocities of 1.5 ~$\rm km ~ s^{-1}$  and dynamical time scales of about 1 minute \citep[][]{vanB2018}. The photospheric foot-point motions generate transverse MHD waves that propagate upward along the magnetic field lines.  The waves are reflected due to the density variations in the photosphere, chromosphere, and corona creating inward propagating waves. The counter-propagating waves interact nonlinearly, resulting in turbulence. The energy is dissipated as a result of turbulence, raising the temperature of the corona to 2$-$3 MK.

\begin{figure}[http]
\epsscale{1.15}
\plotone{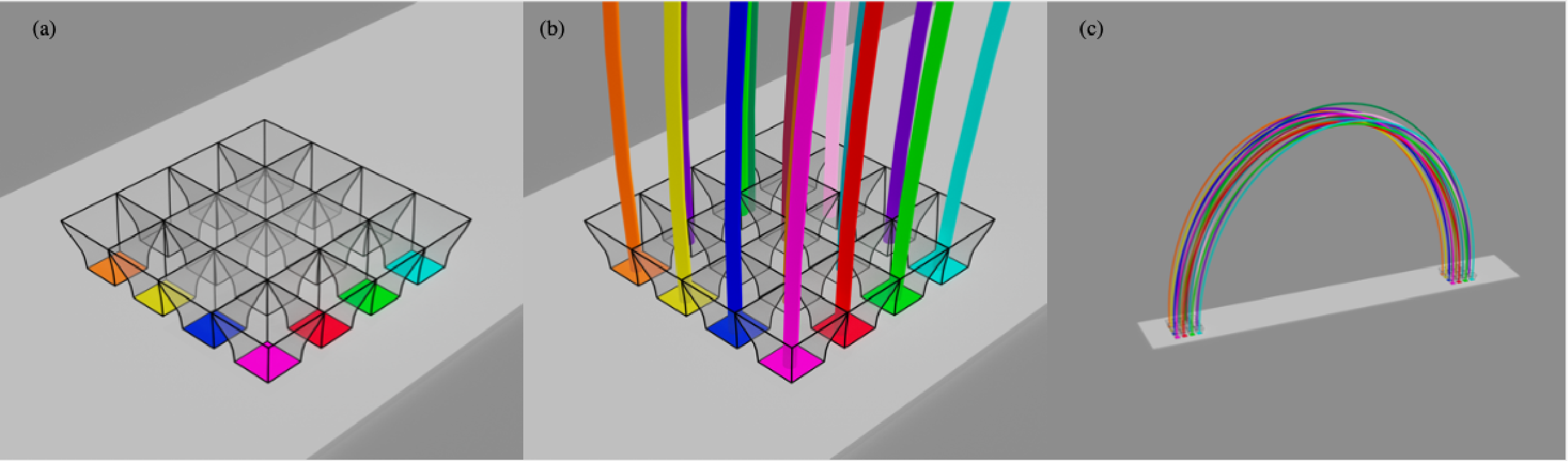}
\caption{(a) An array of $4{\times}4$ flux tubes merging into a space-filling field. (b) Close-up of the left footpoint.  The colored squares indicate the intersections of the flux tubes with the plane $z=0$. (c) Magnetic field lines heated by Alfv\'{e}n wave turbulence (AWT). The colors of the field lines are randomly selected.}
\label{fig:fig3}
\end{figure}

Here, we consider a more advanced version of our Alfv\'{e}n wave turbulence model, consisting of a collection of 16 photospheric flux tubes with square cross sections. This model is the same model that was presented in \citet[][]{vanB2017} and is shown in Figure~\ref{fig:fig3}. We adopt this model to simulate the emission of the active region loop arcade. Here, we provide a brief review of this model, and in the next section, we compare the non-thermal velocity from this model with observations of active regions listed in Table 1.

Figure~\ref{fig:fig3} shows the discrete flux tubes. The random foot point motions imposed at the photosphere generate Alfven waves inside each magnetic field line. The flux tubes expand with height and merge at the height of 520 km in the low chromosphere. The merged field extends from the chromosphere at one end of the loop to the chromosphere at the other end, so the transition regions are located within the merged field (the coronal loop length $L_{\rm c} = 98.4$ Mm). Note that in Figure~\ref{fig:fig3}, the discrete flux tubes are shown and not the merged field lines. We used the RMHD approximation \citep[e.g.,][]{Strauss1976, Strauss1997} to model the waves. The magnetic and velocity perturbations are described by  $\delta {\bf B}_\perp = \nabla h \times {\bf B}_0$ and $\delta {\bf v}_\perp = \nabla f \times \hat{\bf s}$, respectively, where ${\bf B}_0 ({\bf r})$ is the  background field, $\hat{\bf s}$ is the unit vector along ${\bf B}_0$, $h({\bf r},t)$ is the magnetic flux function, and $f({\bf r},t)$ is the velocity stream function. 

 At the merging height the waves can travel from the flux tubes into the merged field or vice versa.  In the RMHD approximation, the magnetic field strength $B_0 (s)$ and plasma density
$\rho (s)$ are assumed to be constant over the cross section both for the flux tubes and for
the merged field. 

The Alfv\'{e}n waves are described in terms of stream
functions $f_{\pm} (x,y,s,t)$ for the Elsasser variables:
\begin{equation}
{\bf z}_{\pm} (x,y,s,t) = \nabla_\perp f_{\pm} \times \hat{\bf s} ,
\end{equation}
where $\nabla_\perp$ is the perpendicular gradient. The velocity
stream function $f = (f_{+} + f_{-})/2$, and the magnetic flux function $h = (f_{-} - f_{+})/
(2 v_{\rm A})$, where $v_{\rm A} (s) \equiv B_0 / \sqrt{4 \pi \rho}$ is the Alfv\'{e}n speed.

The dynamics of the waves are described by 
\begin{equation}
\frac{\partial \omega_{\pm}} {\partial t} = \mp v_{\rm A} \frac{\partial \omega_{\pm} }
{\partial s} + \frac{1}{2} \frac{dv_{\rm A}} {ds} \left( \omega_{+} - \omega_{-} \right)
+ {\cal N}_{\pm} + \tilde{\nu}_{\pm} \nabla_\perp^2 \omega_{\pm} ,   \label{eq:rmhd1}
\end{equation}
where $\omega_{\pm} \equiv - \nabla_\perp^2 f_{\pm}$ are the vorticities of the waves. 
The four terms on the right-hand side of Equation (\ref{eq:rmhd1}) describe the wave propagation,
linear couplings resulting from gradients in Alfv\'{e}n speed, nonlinear coupling between
counter-propagating waves, and wave damping. ${\cal N}_{\pm}$ are nonlinear terms, and $\tilde{\nu}_{\pm}$ are viscosities.
The nonlinear terms are given by
\begin{equation}
{\cal N}_{\pm} = - \onehalf [ \omega_{+} , f_{-} ] - \onehalf [ \omega_{-} , f_{+} ] 
\pm \nabla_\perp^2 \left( \onehalf [ f_{+} , f_{-} ] \right) 
\end{equation}
where $[ \cdots , \cdots ]$ is the bracket operator:
\begin{equation}
[a,b] \equiv \frac{\partial a} {\partial x} \frac{\partial b} {\partial y}
- \frac{\partial a} {\partial y} \frac{\partial b} {\partial x} ,
\label{eq:bracket1}
\end{equation}
with $a(x,y)$ and $b(x,y)$ being two arbitrary functions.
The Fourier analysis is used to describe the dependence
of the waves on the $x$ and $y$ coordinates, and finite-differences in the $s$ direction along
the loop. The dynamics of the waves are simulated for a period of 3000 s during which the Alfv\'{e}n wave turbulence 
is generated along the field lines, depositing energy and heating the corona. 

\section{Effects of Alfv\'{e}n Waves on Spectral Line Broadening}

In this section, we simulate the effects of the Alfv\'{e}n waves on the Doppler shift and Doppler width for the modeled field lines and compare those quantities with the observations of non-thermal velocity presented in Section 2.2 and Appendix A.

In order to compare the current model with observations, we simulate the impact of the modeled waves on the Doppler shift and Doppler width of an observed spectral line, assuming a lateral view of the coronal loop (toward the $+y$ direction). Subsequently, the line-of-sight (LOS) velocity is represented by $v_y$, and the observed Doppler shift is directly proportional to the mean value of $v_y$ along the LOS, denoted as $< v_y >$. Additionally, the non-thermal component of the Doppler width scales with the velocity variance $\sigma_y$, expressed as $\sigma_y^2 = < v_y^2 > - ( < v_y > )^2$. We simplify by assuming constant emissivity and thermal width of the spectral line across the loop cross section. Consequently, $< v_y >$ and $< v_y^2 >$ are approximated as simple averages over the $y$ coordinate in our numerical model. To accommodate instrumental effects, we also average in the $x$ direction over a distance $\Delta x = D_0 \Delta \theta$, where $\Delta \theta$ represents the angular resolution of the instrument, and $D_0$ denotes the Sun-Earth distance.

\begin{figure}
\epsscale{1.00}
%\plotone{figure23.eps}
\plotone{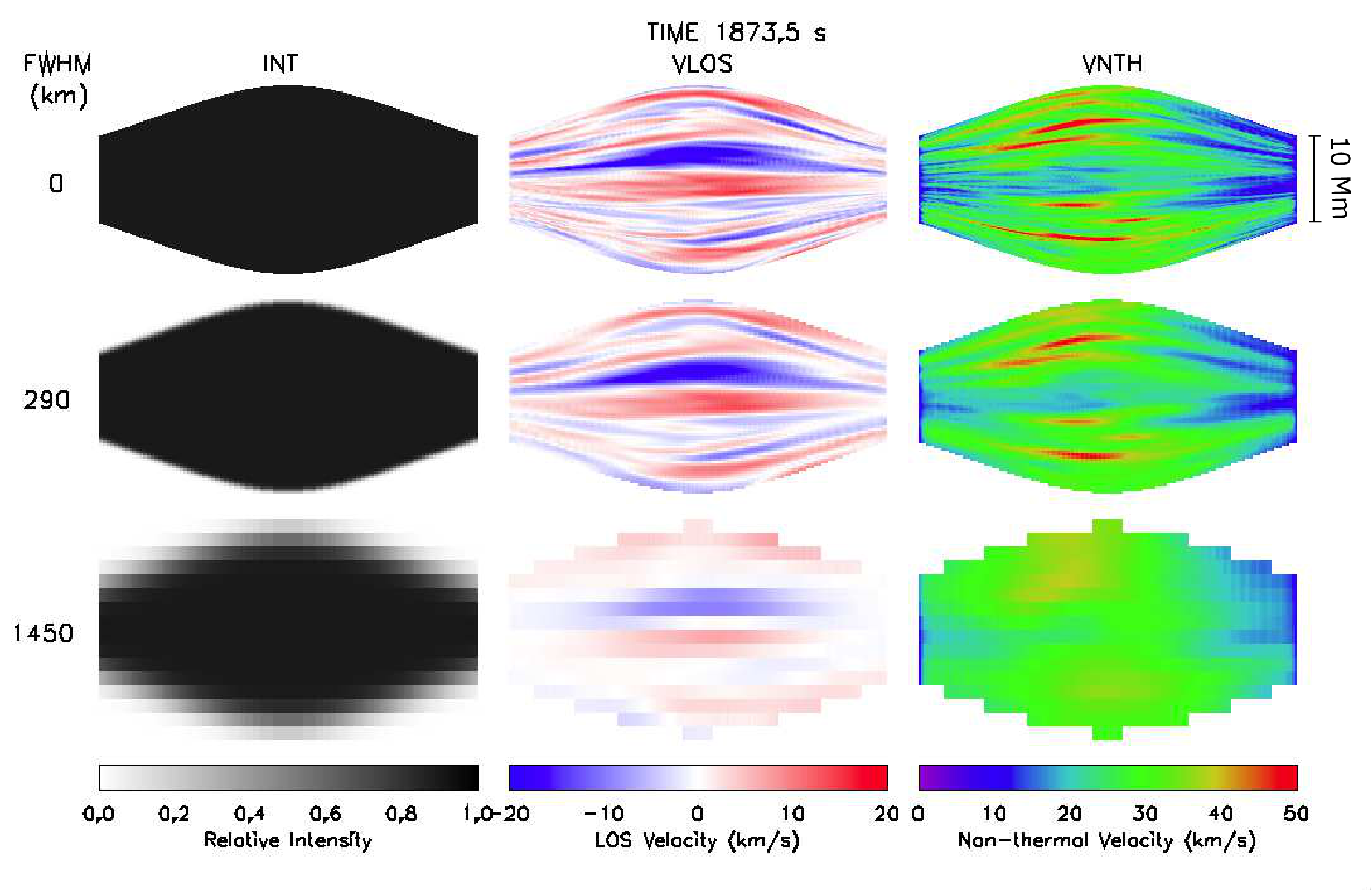}
\caption{Effect of the modeled waves on the Doppler shift and Doppler width of a spectral
line, assuming the merged field lines are viewed from the side. The time in the model is 1873.5~s. The three columns show the intensity (INT), LOS velocity (VLOS), and non-thermal velocity (VNTH). The three rows correspond to
different values of the spatial resolution of our model, MUSE, and EIS ($\Delta x =$ 0, 290 and 1450 km, respectively). 
The field curvatures are neglected, so the loop axes
runs horizontally through the middle of each image. The transverse (vertical) scale is greatly
expanded compared to the longitudinal scale. Spatial smearing and pixelation have been
applied to the data. The velocity amplitudes are indicated by the
color bars.}
\label{fig:fig4}
\end{figure}

Figure~\ref{fig:fig4} shows the Doppler velocities at time $t = 1873.5$~s in the simulation. 
The three columns show maps of intensity (INT), average LOS velocity
(VLOS), and average non-thermal velocity (VNTH). The top row shows the maps from our RMHD model based on Alfv\'en wave
turbulence. The second row shows maps for an instrument with a spatial resolution of
290~km, such as the Multi-slit Solar Explorer \citep[MUSE;][]{DePontieu2021} and pixel size of 121~km, when projected onto the Sun. \citet[]{Brooks2013} found this resolution to be sufficient to resolve coronal loops. The bottom row represents the EIS instrument with a spatial resolution of about 1450 km (2$^{\prime\prime}$) and pixel size of 725 km (1$^{\prime\prime}$).
In all panels of  Figure~\ref{fig:fig4}, we only show the merged field and ignore the field curvature so each image is the loop projected on the $(s,x)$ plane. The width of the image $w(s)$ varies with position $s$ along the loop, but the vertical size of each image corresponds to a width of 10 Mm along the entire length of the loop. We have just expanded the 
vertical scale relative to the horizontal scale to show the velocity structures inside the loop more clearly.
The first column of Figure~\ref{fig:fig4} shows the intensity using an inverted
greyscale. As the color bar indicates, black signifies higher intensity. At full resolution, the boundaries of the observed structure appear sharp due to the assumption of a square cross section for the loop and the emissivity is presumed to diminish beyond the simulation domain. However, at lower resolutions, the edges become less defined, and the pixel size becomes more apparent. This figure is similar to Figure~\ref{fig:fig4}  presented in \citet[][]{vanB2017}, except that 
those authors looked at these maps at a different time in their simulation.

The middle column shows the predicted VLOS as a color-scale image. 
The velocity scale is given at the bottom of the column. The upper panel
shows the simulated velocity map from our model. Note that VLOS varies rapidly in the
$x$ direction (vertical) while showing gradual changes in the $s$ direction (horizontal), suggesting that the internal
motions exhibit coherence along the loop. The changes in the VLOS appear on a length scale of about 500 km. The rms value of VLOS over the entire image is 5.8 $\rm km ~ s^{-1}$. The VLOS in the middle panel is shown for an instrument such as MUSE.
The velocity pattern from this instrument is almost identical to the velocities from our model in the top panel.
 The rms velocity from this instrument is 5.1 $\rm km ~ s^{-1}$, suggesting such instruments would resolve well the velocity variations from our model. 
 
 The bottom VLOS panel shows the velocity fluctuations with an rms value of 2.7 $\rm km ~ s^{-1}$ from an instrument with a spatial resolution of EIS. The uncertainty in EIS velocity measurements is on the order of 4.5 $\rm km ~ s^{-1}$ \citep[][]{Kamio2010}, though higher sensitivity can be inferred from time-series spectra \citep[][]{Mariska2010}. Still, detecting the VLOS fluctuations simulated here would be challenging with EIS.
 
The right column of Figure~\ref{fig:fig4} shows the predicted non-thermal velocity (VNTH) from our model, a MUSE-like instrument, and the EIS instrument. 
The corresponding velocity scale is given at the bottom of this column. The value of VNTH
averaged over the image from the transverse waves in our model is about 27 $\rm km ~ s^{-1}$. This is consistent 
with the spectroscopic observations of non-thermal velocities from EIS shown in the bottom right panels of Figures~1, 2, and A1$-$A20. These velocities have a range of $\sim$ 17$-$30 $\rm km ~ s^{-1}$. 

\begin{figure}
\epsscale{1.00}
\plotone{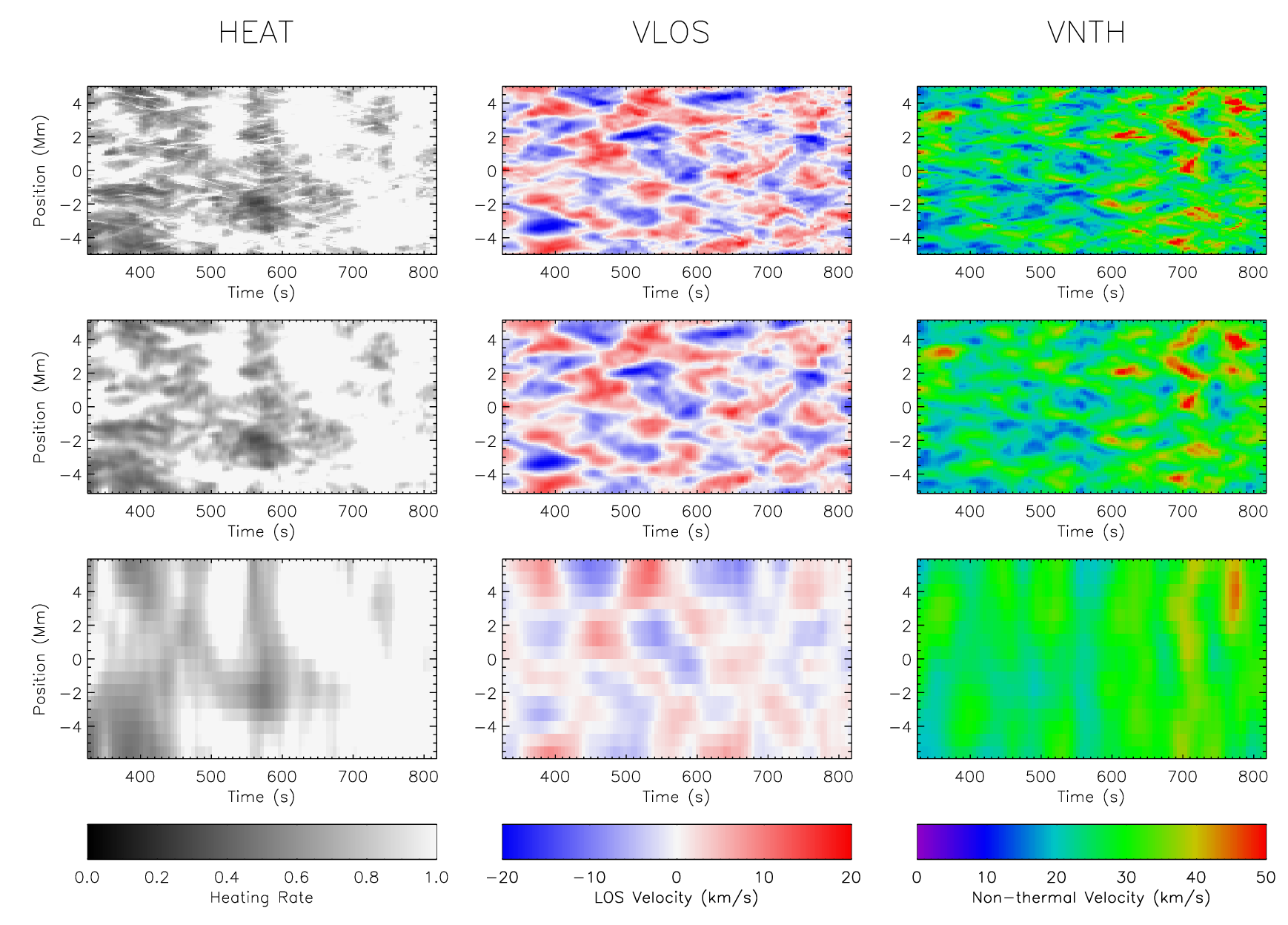}
\caption{Time slices of heating rate (HEAT), LOS Velocity (VLOS) and Non-thermal Velocity (VNTH) at a spatial resolution of our RMHD model (top row), a MUSE-like instrument at 290  km (middle row), and EIS at 1450 km (bottom row).}
\label{fig:fig5}
\end{figure}

Figure~\ref{fig:fig5} shows the time-position plot across the 10 Mm width at the top of the loop. The first column shows the heating rate integrated along the line of sight, the second column the VLOS, and the third column VNTH. The first row is the Alfv\'{e}n wave turbulence model, the second row is the MUSE-like instrument, and the third is the EIS instrument. The non-thermal velocity map from our model shows a variation of $\sim$ 15$-$37 $\rm km ~ s^{-1}$,  in agreement with the VNTH from the analyzed EIS observations (Figures~1, 2, and A1$-$A20) and in accordance with the values shown in the first panel of Figure~\ref{fig:fig4}. The brightenings in the heating rate map from our model (top panel) are indicators of the large heating rates.

\clearpage
\section{Discussion and Conclusion}

In this paper, we presented a review of non-thermal velocity and instrumental width measurements. 
We then analyzed the EIS observations of spectral line broadening in Fe {\sc xvi} 262.984 {\AA}, Fe {\sc xiv} 264.787 {\AA}, Fe {\sc xiv} 270.519 {\AA}, 
Fe {\sc xiv} 274.203 {\AA}, Fe {\sc xv} 284.160 {\AA}, and studied 11 active regions observed with 1$^{\prime\prime}$ and 2$^{\prime\prime}$ resolution slits. We found
that the non-thermal velocities fall in the range of 17--30\,km s$^{-1}$, with an uncertainty of 4.7\,km s$^{-1}$. We also noticed a possible tendency for the non-thermal velocity
to increase with wavelength and/or temperature, although the change is close to the measurement uncertainty.
The non-thermal velocity measurements from the observations were compared with our Alfv\'{e}n wave turbulence model built of multiple flux tubes. 

\citet[][]{Imada2009} analyzed ion thermal temperatures in an active region observed by EIS using the spectral lines 
Fe {\sc xvi} 262.98 {\AA} and S {\sc xiii} 256.69 {\AA}. They found that the typical non-thermal velocities were 13 $\rm km ~ s^{-1}$ at temperatures of $\sim2.5$ 
MK. The highest non-thermal velocities ($>20$ $\rm km ~ s^{-1}$) were observed between the bright points in Fe {\sc xvi}. 
 \citet[][]{Testa2016} analyzed spectroscopic observations of the Fe {\sc xii} emission line at 1349.4 {\AA} 
using the Interface Region Imaging Spectrograph  \citep[IRIS;][]{DePontieu2014} and at 
195.119 {\AA} using EIS observations. IRIS has a high spatial resolution of $0.33^{\prime \prime}$ and observes the chromosphere and 
transition region, creating slit-jaw images and high resolution spectra. In their study of the spectral properties of 
Fe {\sc xii}, \citet[][]{Testa2016} considered two active region data sets where the spectral properties were determined by fitting 
the spectra with one Gaussian in the case of IRIS observation and two Gaussians in the case of EIS observations. The instrumental 
width in their calculation for IRIS was estimated to be of the order of 4 $\rm km ~ s^{-1}$ \citep[][]{DePontieu2014} 
while for EIS it was of the order of 61$-$75  $\rm km ~ s^{-1}$. The non-thermal width measured from IRIS data 
sets was 10$-$15  $\rm km ~ s^{-1}$ for active region moss while it peaked at $\sim$ 20 $\rm km ~ s^{-1}$ from EIS observations. 
\citet[][]{Brooks2016} found that the absolute calibration of the EIS data could potentially increase the line width, and
this was confirmed by \citet[][]{Testa2016}.  Both \citet[][]{Testa2016} and 
 \citet[][]{Brooks2016} suggest that  the non-thermal velocity does not increase with temperature. 

Note, however, that differing assumptions and observational targets make direct comparison of our results with previous work difficult. Our measurements focus on the non-thermal velocities of whole active regions, whereas \citet[][]{Imada2009} studied the spatial variation of non-thermal velocities within an active region. They
also assumed that Fe {\sc xvi} 262.98 {\AA} and S {\sc xiii} 256.69 {\AA} have the same ion temperature and non-thermal velocity, whereas we assume the lines are formed
at the temperature of the peak of the ionization fraction, which is slightly higher for Fe {\sc xvi} 262.98 {\AA}. \citet[][]{Testa2016} made measurements
in the moss at the footpoints of high temperature loops, and \citet[][]{Brooks2016} studied isolated portions of the high temperature loop tops. It is possible that
non-thermal velocities behave differently with temperature in different specific features of an active region, and this needs further investigation.
 
In our study, we analyzed the average non-thermal velocity for 11 active regions and found it to be in the range of 17$-$30 $\rm km ~ s^{-1}$. We compared these observations with the Alfv\'{e}n wave turbulence model of 16 interacting field lines.  The Doppler shift and Doppler width of the simulated waves in these loops presented in Figure~\ref{fig:fig4} shows that the rms value of VLOS over the entire velocity map is 5.8 $\rm km ~ s^{-1}$. This VLOS rms could be resolved by an instrument such as MUSE \citep[][]{DePontieu2021} with its high spatial resolution of 290 km. This confirms that these velocity variations would be well resolved in such observations. The simulated non-thermal velocity from Figure~\ref{fig:fig5} showed these velocities to have a range of  15$-$37  $\rm km ~ s^{-1}$, consistent with the non-thermal velocities from the EIS observations.  

\acknowledgements 
M. Asgari-Targhi,  M. Hahn, and D. W. Savin are supported under contract NNM07AB07C from NASA to the Smithsonian Astrophysical Observatory (SAO) and Columbia University. The work of D.H. Brooks was performed under contract to the Naval Research Laboratory and was funded by the NASA Hinode program.

\clearpage

\appendix
%\section{Appendix}
\renewcommand{\thefigure}{A\arabic{figure}} % Redefine figure numbering to include "A" for appendix
\setcounter{figure}{0} % Reset the figure counter to start from 1 in the appendix
\section{Observational Results: Gaussian Analysis}

\subsection{Active Regions Observed With 1$^{\prime\prime}$ Slit}
\label{1arcslit}

\begin{figure}
\epsscale{0.65}
\plotone{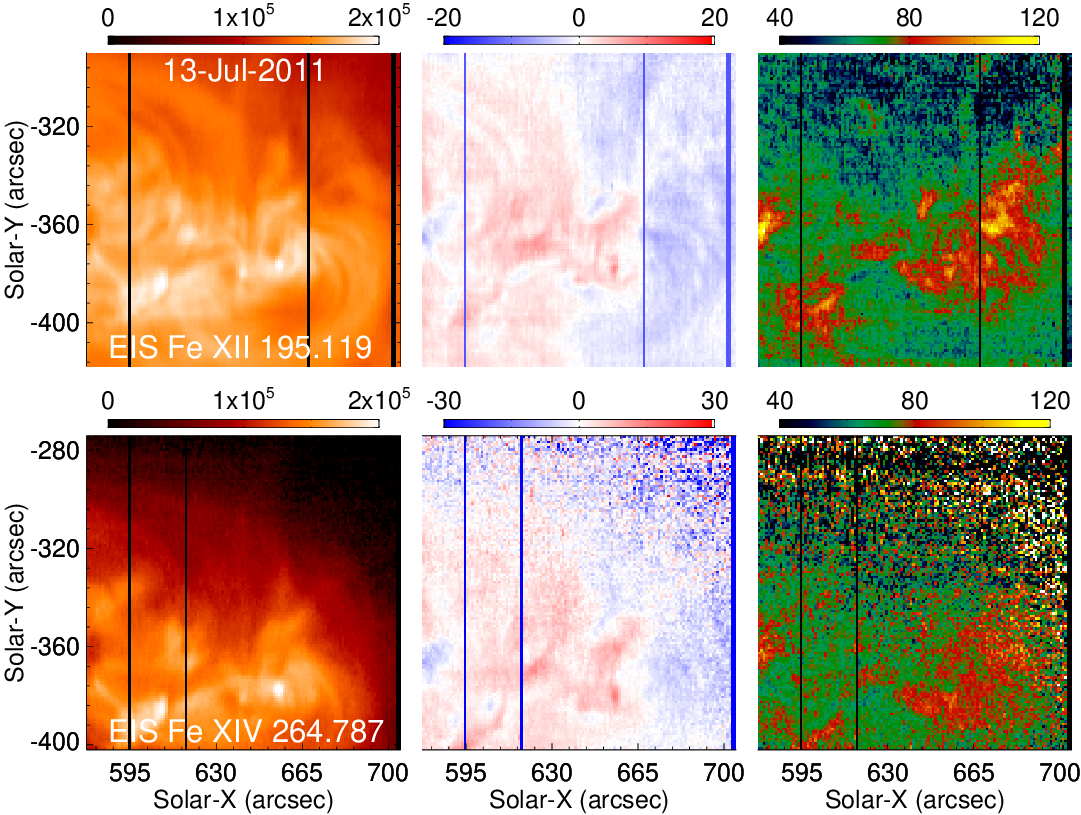}
\caption{Same as Figure 2 but for the active region NOAA 11247 observed on 2011 July 13.}
\label{fig:fig6}
\end{figure}

\begin{figure}[http]
\epsscale{1.00}
\plotone{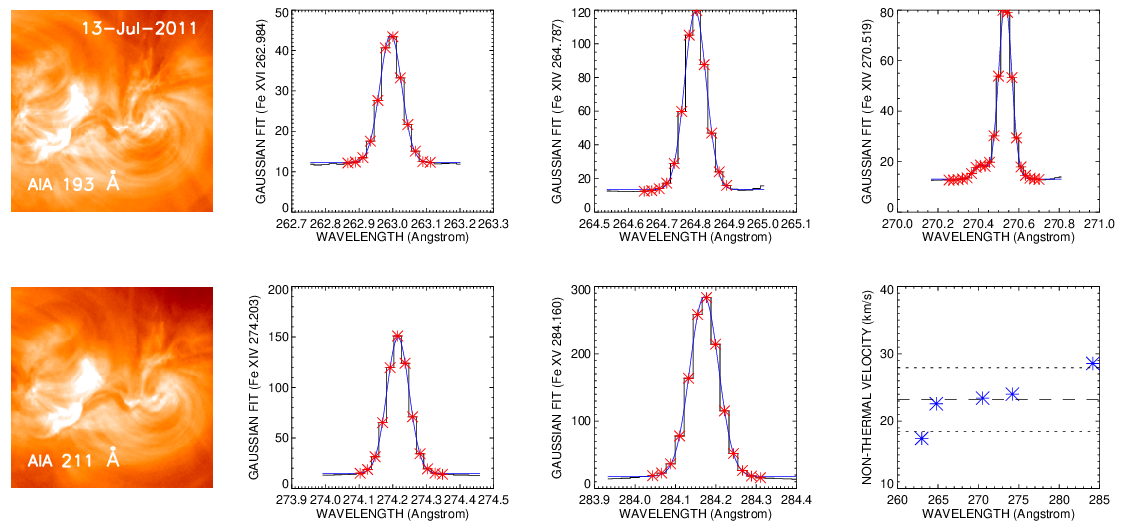}
\caption{Same as Figure 1 but for the active region NOAA 11247 observed on 2011 July 13.}
\label{fig:fig7}
\end{figure}

\begin{figure}[http]
\epsscale{0.65}
\plotone{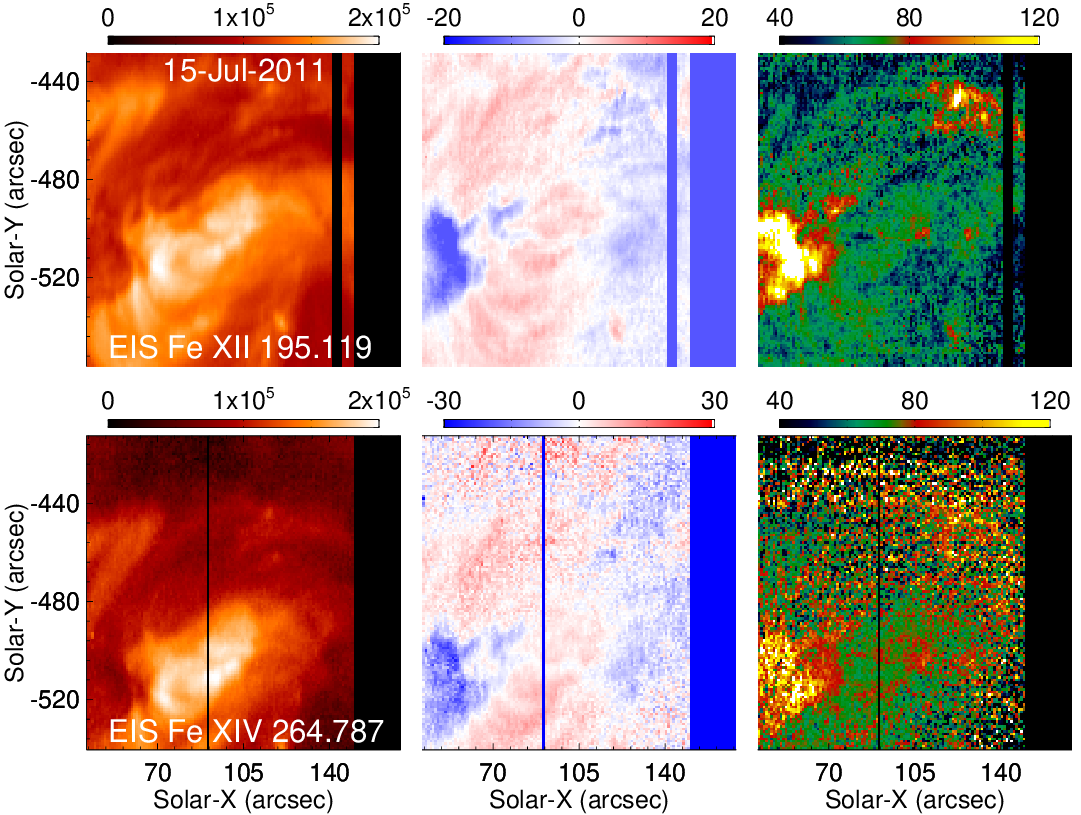}
\caption{Same as Figure 2 but for the active region NOAA  11250 observed on 2011 July 15.}
\label{fig:fig8}
\end{figure}

\begin{figure}[http]
\epsscale{1.00}
\plotone{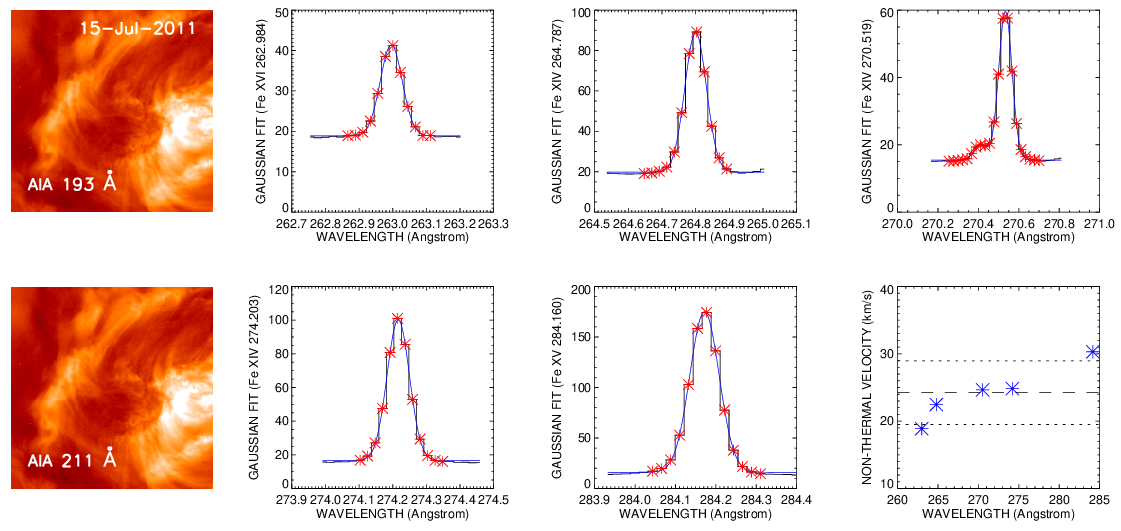}
\caption{Same as Figure 1 but for the active region NOAA 11250 observed on 2011 July 15.}
\label{fig:fig9}
\end{figure}

Figures~\ref{fig:fig6} $-$\ref{fig:fig9} show our results using the 1$^{\prime\prime}$ slit.  These results are in good agreement with those from the 2$^{\prime\prime}$ slit. We estimated the intensity, Doppler velocity, and line width by a single Gaussian fitting. We used the  eis\_auto\_fit routine to fit a Gaussian to the spectrum. All the dark stripes seen in the images were set to the missing values in the data window structure and they are ignored by the fitting routine. We used eis\_get\_fitdata to extract the line quantities from the output structure of the eis\_auto\_fit. 

Figures~\ref{fig:fig6} and \ref{fig:fig8} show the EIS intensity, velocity, and line width for the active regions 11247 and 11250 from Table 1 observed with $1^{\prime\prime}$ slit. The first column in Figures \ref{fig:fig7} and \ref{fig:fig9} shows intensity maps from EIS or AIA. The other panels show the Gaussian fit  to the spectral  lines for Fe {\sc xvi} 262.984  {\AA}, Fe {\sc xiv} 264.787 {\AA}, Fe {\sc xiv} 270.519 {\AA}, Fe {\sc xiv} 274.203  {\AA}, and Fe {\sc xv} 284.160 {\AA}.  The last panel in the figures plots the FWHM as a function of wavelength. The line profiles are averages over the whole active region.
                                                                                                                                                                                                                                                                                                                                                                                                                                                                                                                                                                                                                                                                                                                                                                                                                                                                                                                                                                                                                                                                                                                              
Figure~\ref{fig:fig7} presents the active region 11247 observed on 2011 July 13. The non-thermal velocity is 18 $\rm km ~ s^{-1}$ for Fe {\sc xvi} 262.984 {\AA} and has a steady increase to 31 $\rm km ~ s^{-1}$ for Fe {\sc xv} 284.160 {\AA}. The non-thermal line broadening is similar for Fe {\sc xiv} 270.519 {\AA} and Fe {\sc xiv} 274.203  {\AA}.

Figure~\ref{fig:fig9} shows the active region 11250 observed on 2011 July 15. The non-thermal velocity for this region has the same trend as 
the active regions shown in Figures~\ref{fig:fig1}, \ref{fig:fig7}, \ref{fig:fig11}, \ref{fig:fig19}, \ref{fig:fig21}, and \ref{fig:fig25}.
In all these regions, the velocity increases between the wavelengths 262.984 {\AA} and 284.160  {\AA}, reaching its highest value at 284.160  {\AA}.

\subsection{Observations With 2$^{\prime\prime}$ Slit}
\begin{figure}[http]
\epsscale{0.65}
\plotone{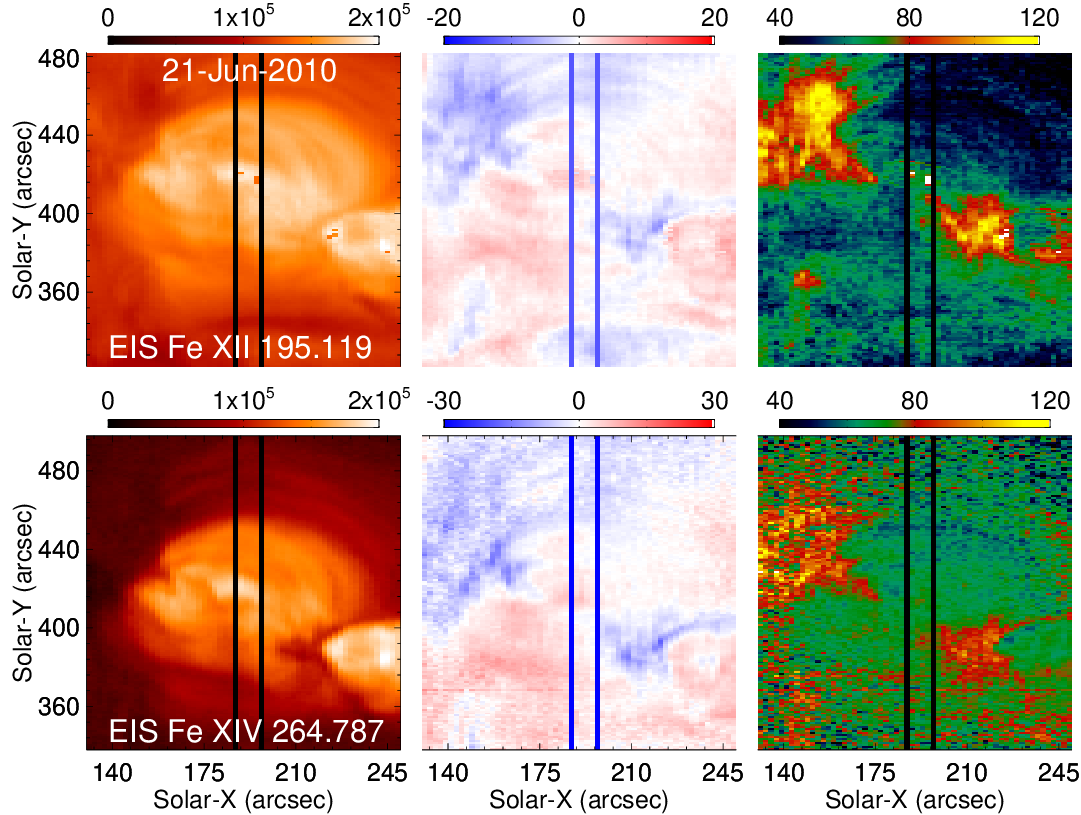}
\caption{Same as Figure 2 but for the active region NOAA 11082 observed on 2010 June 21.}
\label{fig:fig10}
\end{figure}

\begin{figure}[http]
\epsscale{1.00}
\plotone{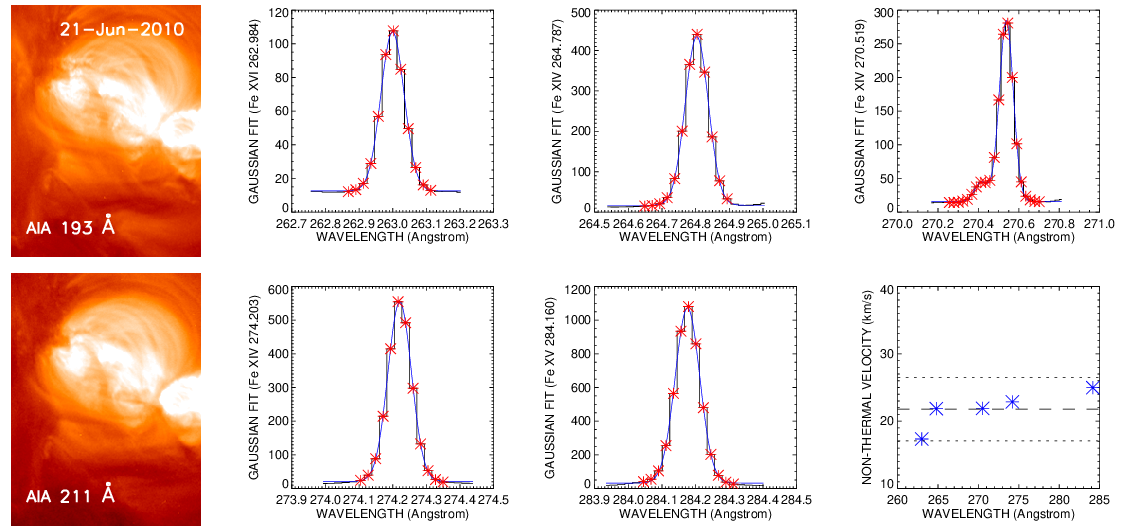}
\caption{Same as Figure 1 but for the active region NOAA 11082 observed on 2010 June 21.}
\label{fig:fig11}
\end{figure}

\begin{figure}[http]
\epsscale{0.65}
\plotone{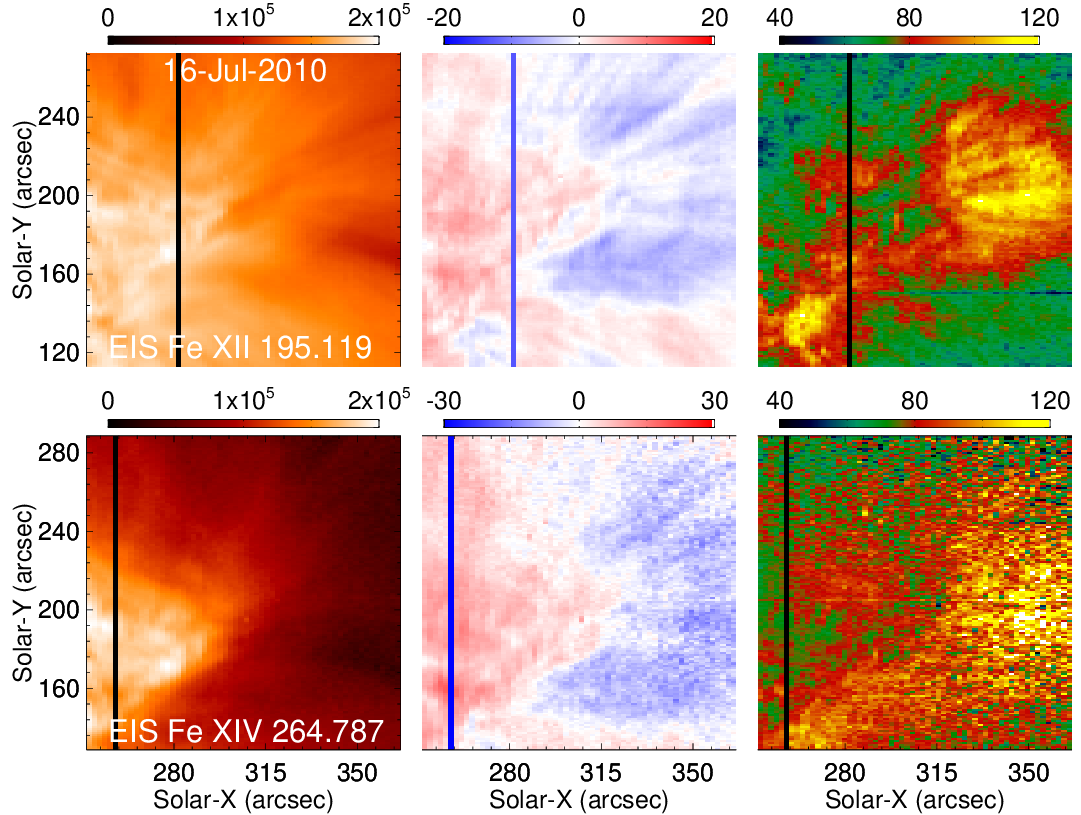}
\caption{Same as Figure 2 but for the active region NOAA 11087 observed on 2010 July 16.}
\label{fig:fig12}
\end{figure}

\begin{figure}[http]
\epsscale{1.00}
\plotone{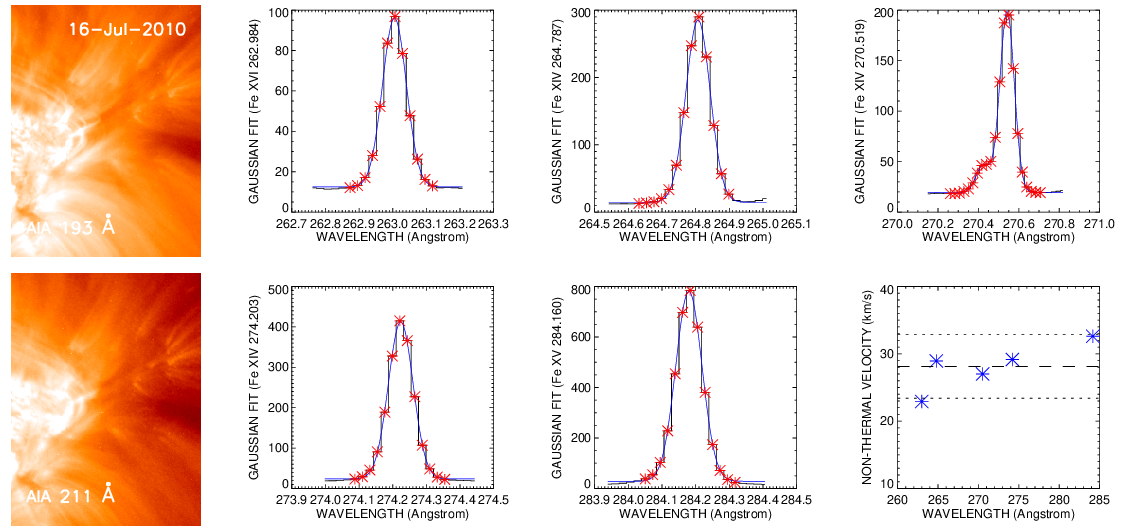}
\caption{Same as Figure 1 but for the active region NOAA 11087 observed on 2010 July 16.}
\label{fig:fig13}
\end{figure}

\begin{figure}[http]
\epsscale{0.65}
\plotone{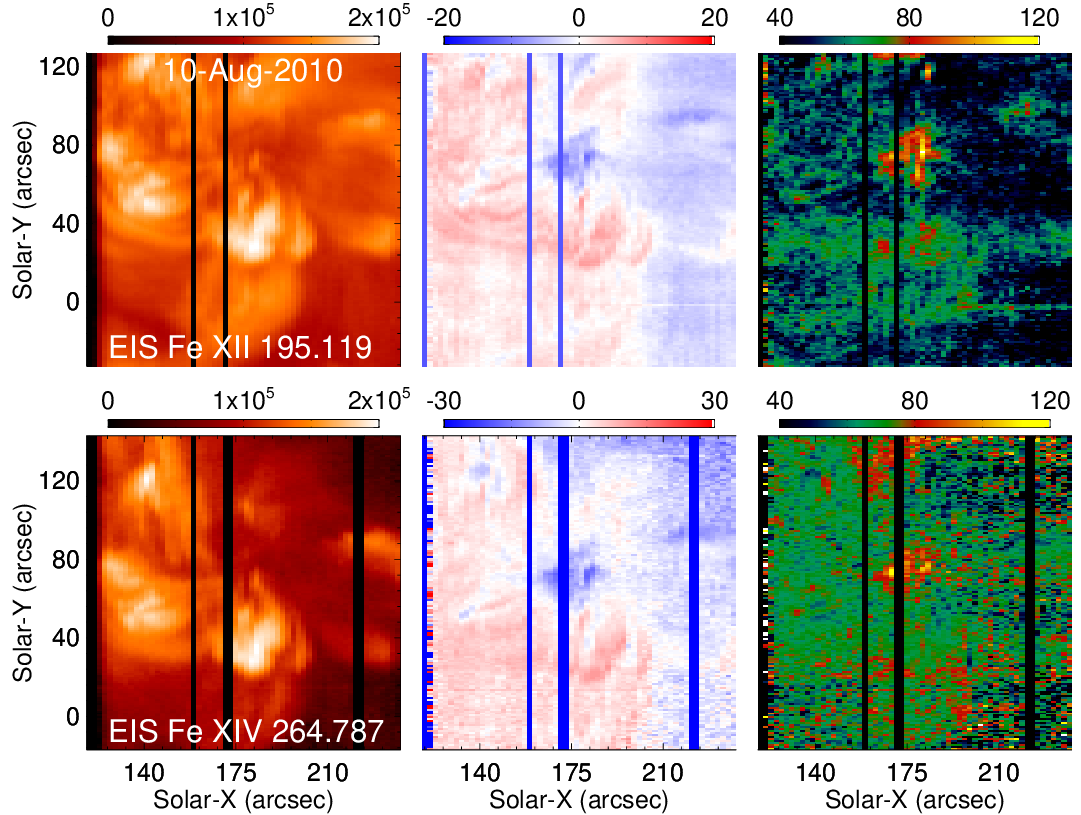}
\caption{Same as Figure 2 but for the active region NOAA 11093 observed on 2010 August 10.}
\label{fig:fig14}
\end{figure}

\begin{figure}[http]
\epsscale{1.00}
\plotone{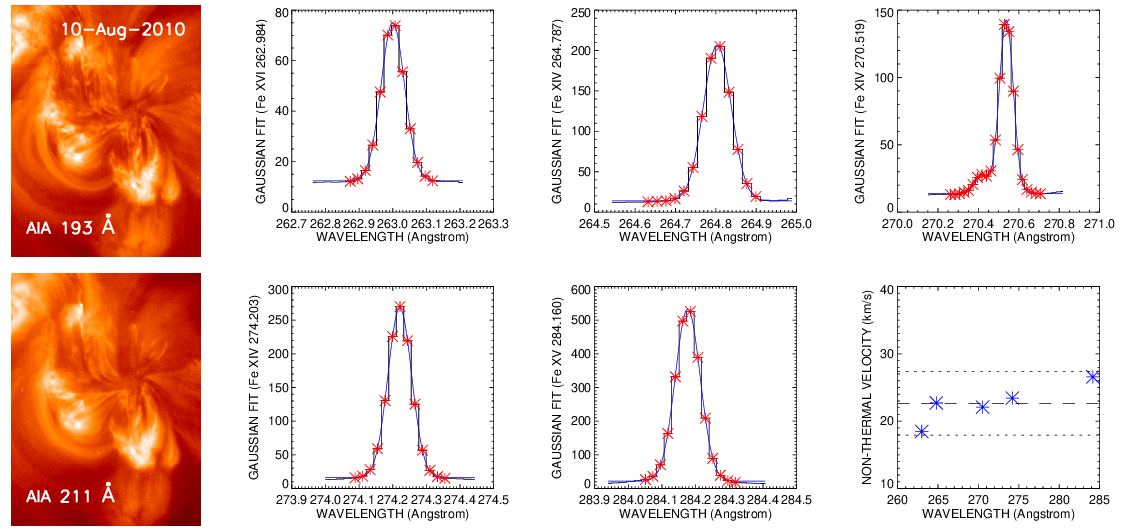}
\caption{Same as Figure 1 but for the active region NOAA 11093 observed on 2010 August 10.}
\label{fig:fig15}
\end{figure}

\begin{figure}[http]
\epsscale{0.65}
\plotone{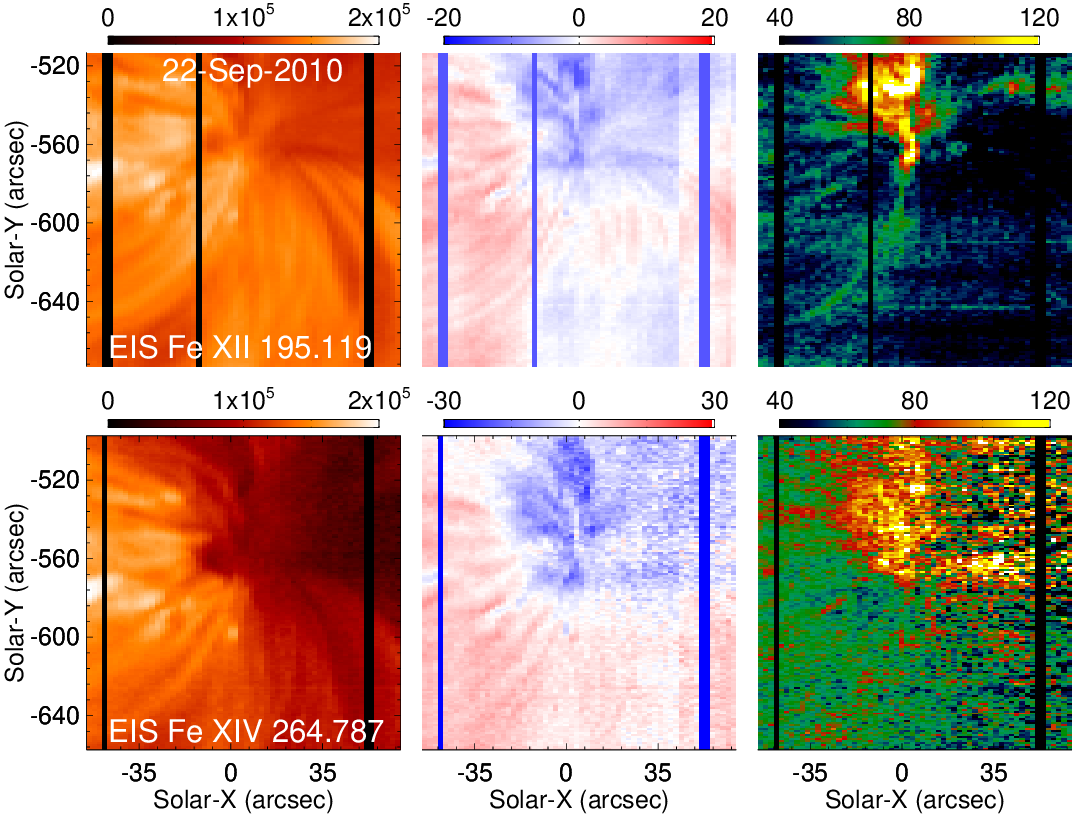}
\caption{Same as Figure 2 but for the active region NOAA 11108 observed on 2010 September 22.}
\label{fig:fig16}
\end{figure}

\begin{figure}[http]
\epsscale{1.00}
\plotone{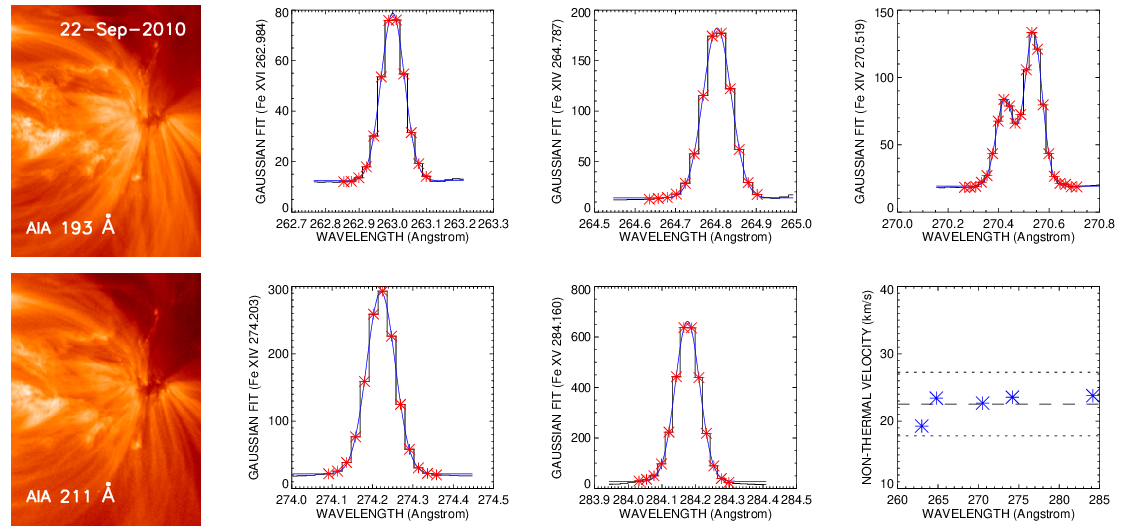}
\caption{Same as Figure 1 but for the active region NOAA 11108 observed on 2010 September 22.}
\label{fig:fig17}
\end{figure}

\begin{figure}[http]
\epsscale{0.65}
\plotone{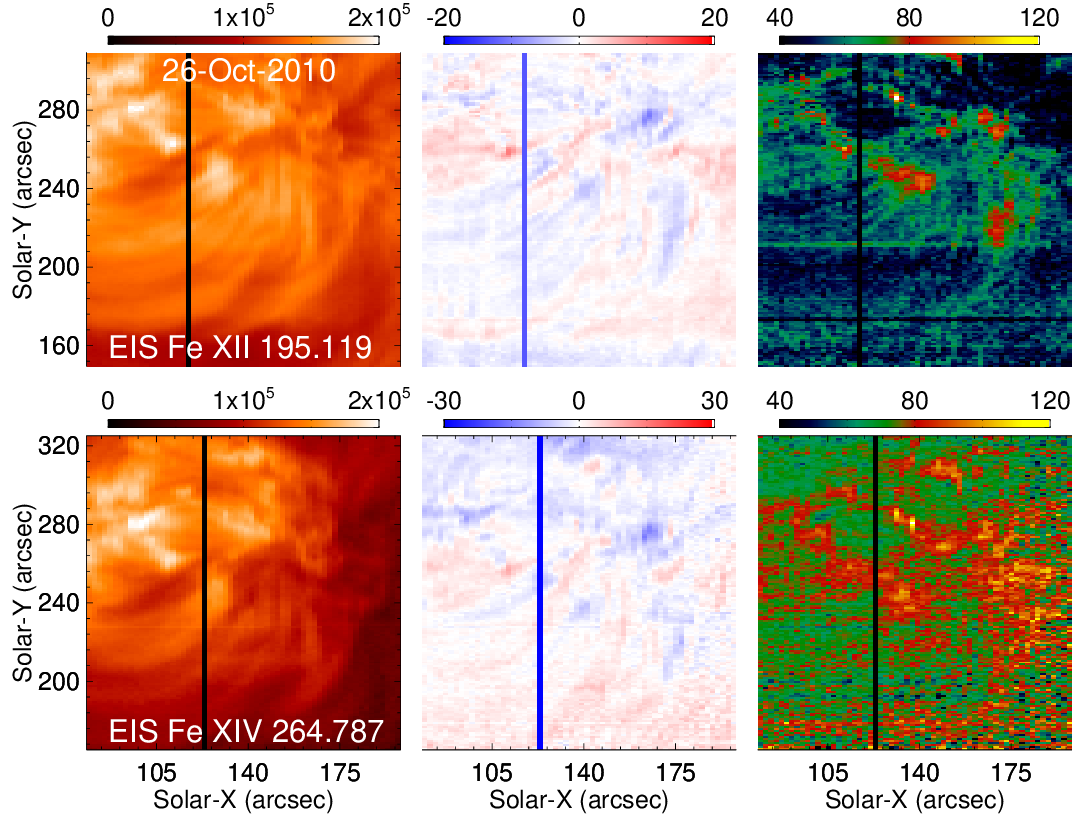}
\caption{Same as Figure 2 but for the active region NOAA 11117 observed  on 2010 October 26.}
\label{fig:fig18}
\end{figure}

\begin{figure}[http]
\epsscale{1.00}
\plotone{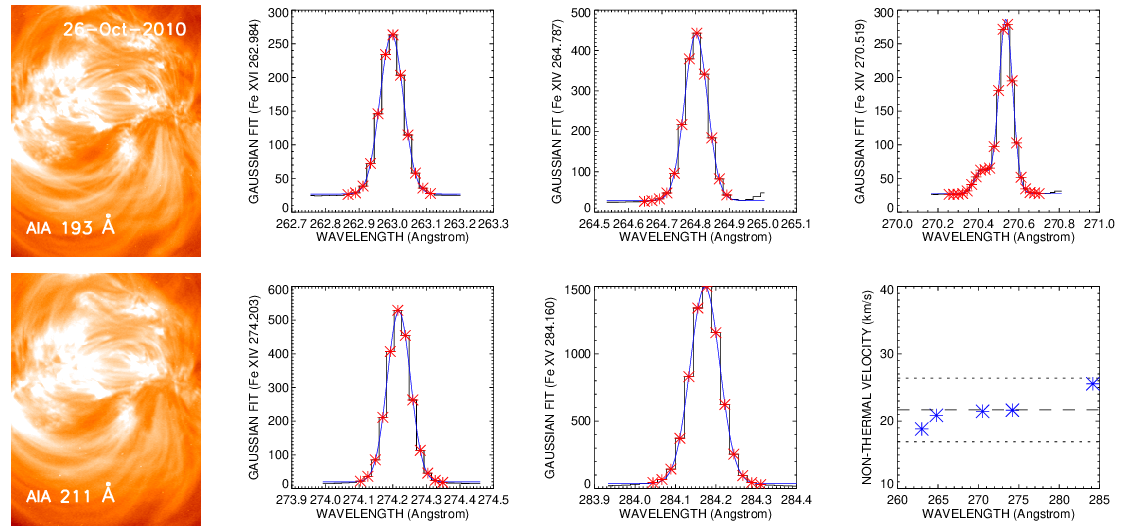}
\caption{Same as Figure 1 but for the active region NOAA 11117 observed on 2010 October 26.}
\label{fig:fig19}
\end{figure}

\begin{figure}[http]
\epsscale{0.65}
\plotone{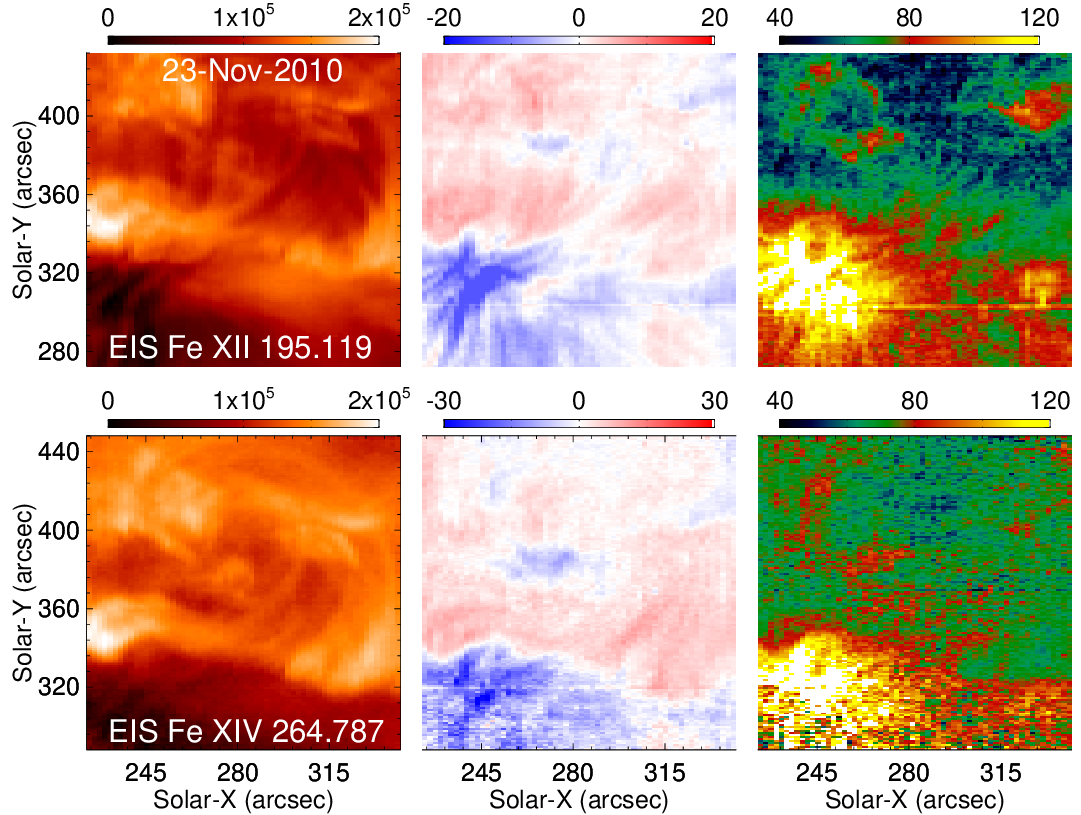}
\caption{Same as Figure 2 but for the active region NOAA 11127 observed on 2010 November 23.}
\label{fig:fig20}
\end{figure}

\begin{figure}[http]
\epsscale{1.00}
\plotone{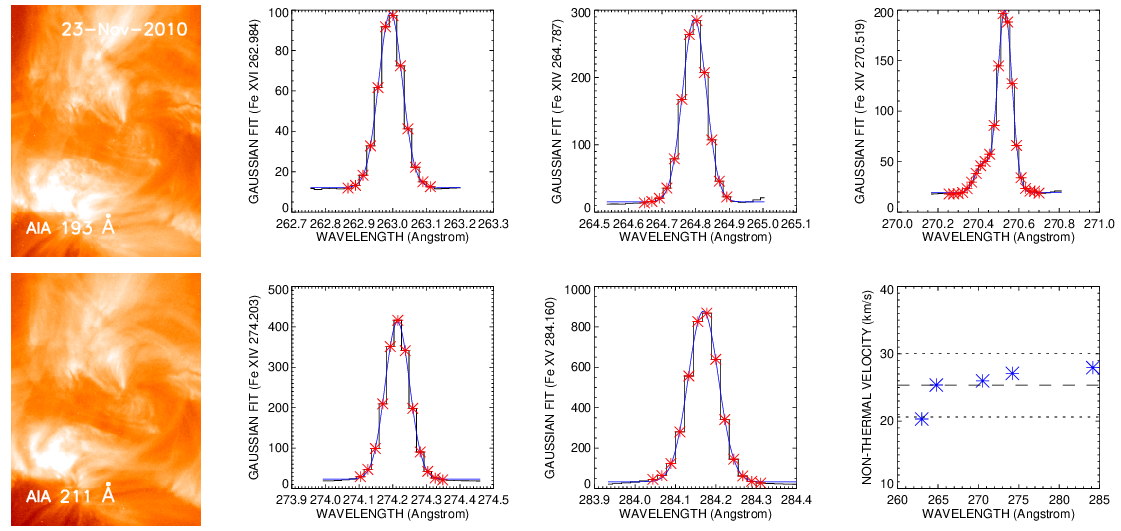}
\caption{Same as Figure 1 but for the active region NOAA 11127 observed on 2010 November 23.}
\label{fig:fig21}
\end{figure}

\begin{figure}[http]
\epsscale{0.65}
\plotone{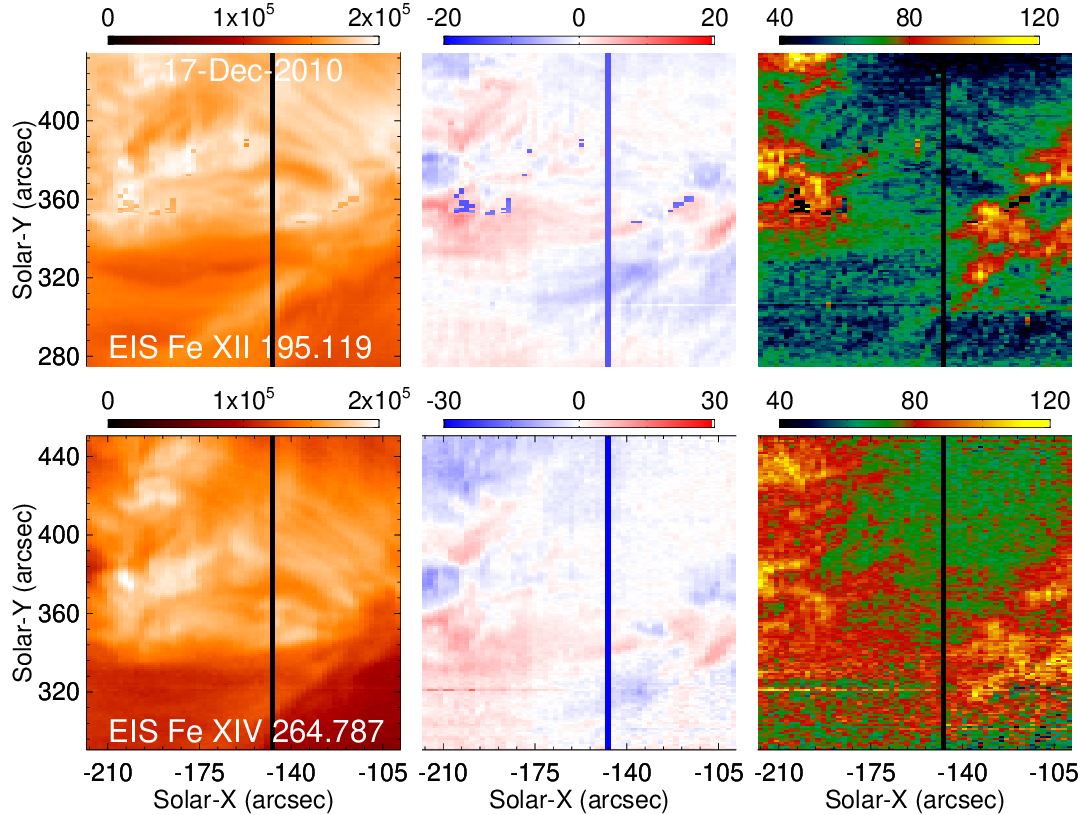}
\caption{Same as Figure 2 but for the active region NOAA 11135 observed on 2010 December 17.}
\label{fig:fig22}
\end{figure}

\begin{figure}[http]
\epsscale{1.00}
\plotone{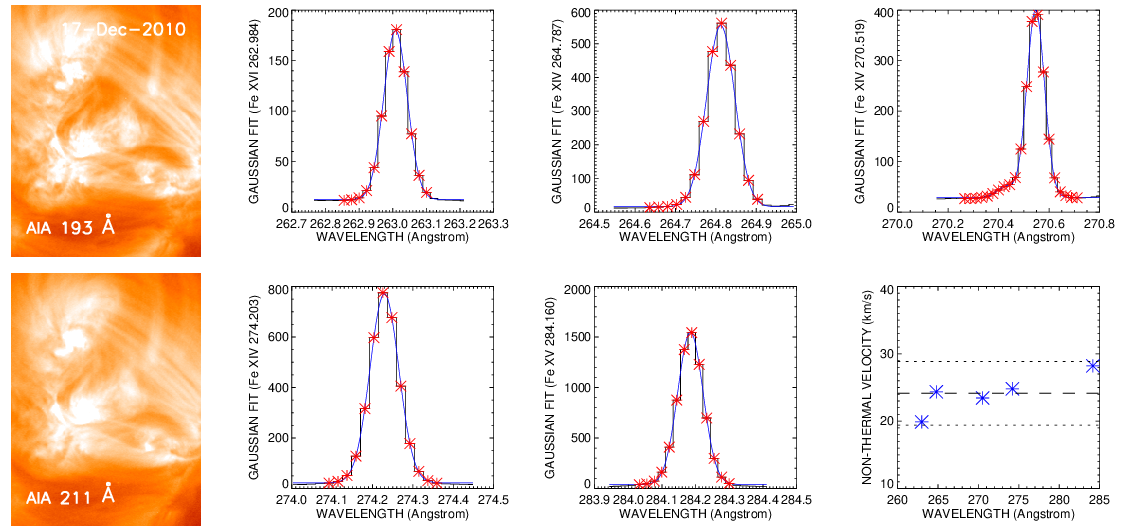}
\caption{Same as Figure 1 but for the active region NOAA 11135 observed on 2010 December 17.}
\label{fig:fig23}
\end{figure}

\begin{figure}[http]
\epsscale{0.7}
\plotone{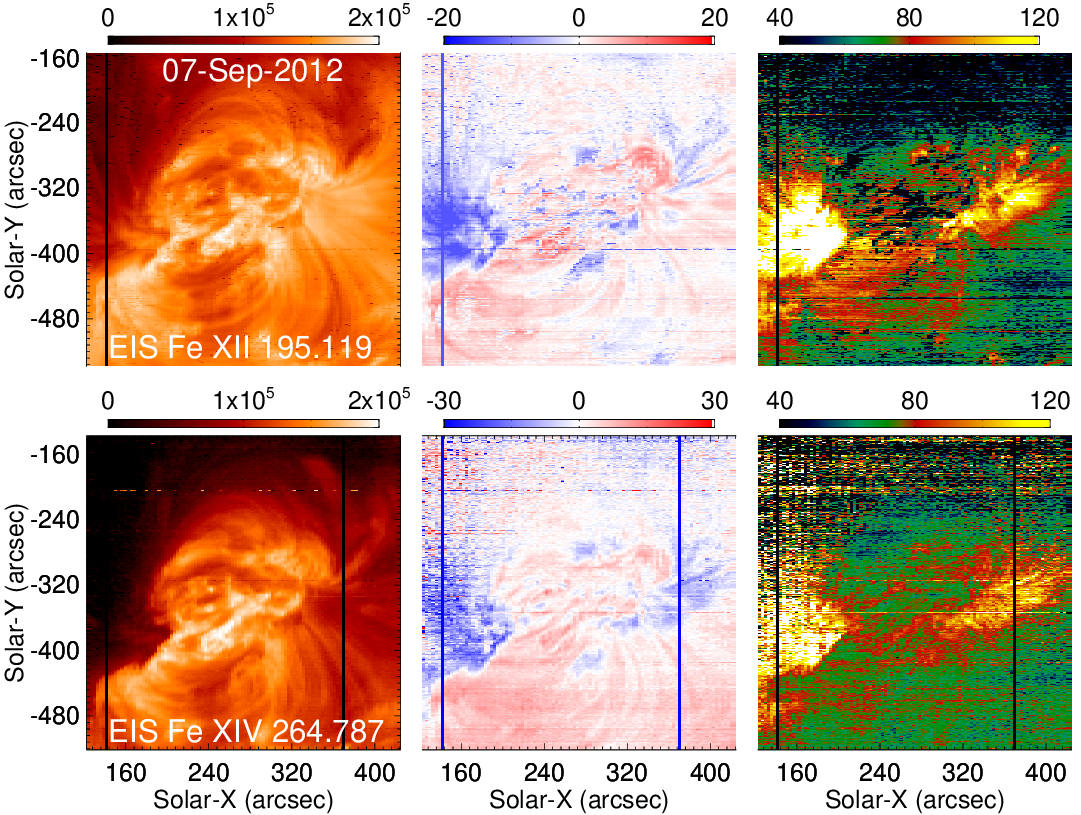}
\caption{Same as Figure 2 but for the active region NOAA 11564 observed on 2012 September 7.}
\label{fig:fig24}
\end{figure}

\begin{figure}[http]
\epsscale{1.00}
\plotone{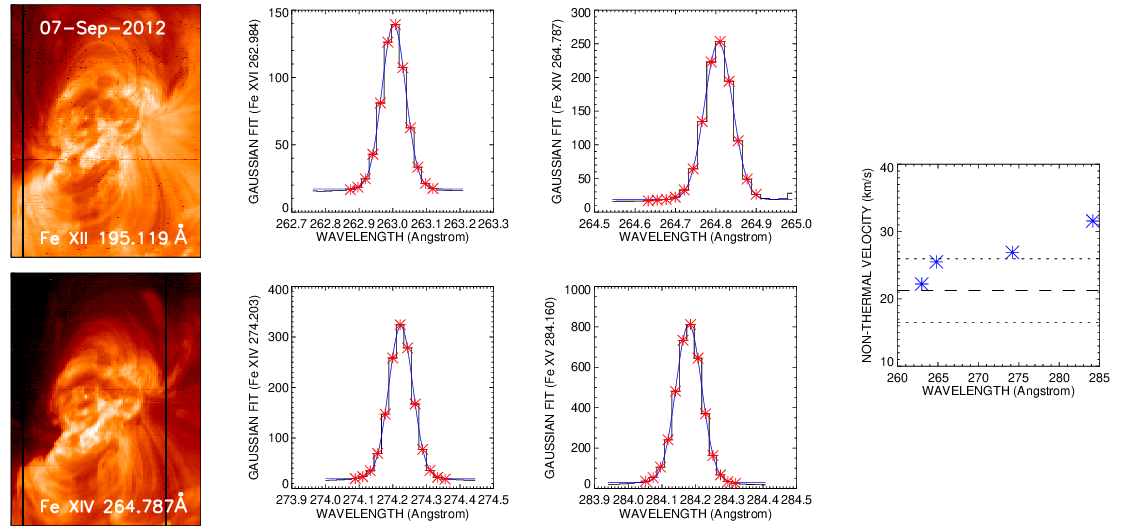}
\caption{Same as Figure 1 but for the active region NOAA 11564 observed on 2012 September 7.}
\label{fig:fig25}
\end{figure}

Figures~\ref{fig:fig10}, \ref{fig:fig12}, \ref{fig:fig14}, \ref{fig:fig16}, \ref{fig:fig18}, \ref{fig:fig20}, \ref{fig:fig22}, and \ref{fig:fig24} show the intensity, velocity, and line width for active regions observed by the 2-arc second slit. 
The first column in Figures~\ref{fig:fig11}, \ref{fig:fig13}, \ref{fig:fig15}, \ref{fig:fig17}, \ref{fig:fig19}, \ref{fig:fig21}, \ref{fig:fig23}, and \ref{fig:fig25} shows context intensity maps from EIS or AIA. Columns 2, 3, and 4 show the Gaussian fit  to the spectral  lines for Fe {\sc xvi} 262.984  {\AA}, Fe {\sc xiv} 264.787 {\AA}, Fe {\sc xiv} 270.519 {\AA}, Fe {\sc xiv} 274.203  {\AA}, and Fe {\sc xv} 284.160 {\AA}. Except for Figure~\ref{fig:fig25}, which does not have the plot for Fe {\sc xiv} 270.519 {\AA} due to lack of proper observation in this line for 2012 September 07.  
\label{2arcslit}

Figure~\ref{fig:fig13} shows the active region 11087 observed on 2010 July 16. The first panel shows the AIA intensity maps in 
193 {\AA} and 211 {\AA}. The non-thermal line broadening increases between 262.984 {\AA} 
and 264.787 {\AA}. It decreases at 270.519 {\AA}, and increases again, reaching a maximum at 284.160  {\AA}. The active regions shown in Figures \ref{fig:fig15}, \ref{fig:fig17}, and \ref{fig:fig23} have the same trend. 
\clearpage

\newpage

\end{document}